\documentclass[prb,twocolumn]{revtex4-1} 

\usepackage{comment}
\usepackage{amsmath} 

\usepackage{graphicx} 

\begin{document}

\title{Photonic Rutherford Scattering: A Classical and Quantum Mechanical Analogy in Ray- and Wave-Optics}
\author{Markus Selmke, Frank Cichos}
\affiliation{Molecular Nanophotonics, Institute of Experimental Physics I, University of Leipzig, 04103, Leipzig}
\email{cichos@physik.uni-leipzig.de} 
\date{\today}

\begin{abstract}
Using Fermat's least optical path principle the family of ray-trajectories through a special but common type of a gradient refractive index lens, $n\left(r\right)=n_0+\Delta n R/r$, is solved analytically. The solution, i.e. the ray-equation $r\left(\phi\right)$, is shown to be closely related to the famous Rutherford scattering and therefore termed photonic Rutherford scattering. It is shown that not only do these classical limits correspond, but also the wave-mechanical pictures coincide: The time-independent Schr\"odingier equation and the inhomogeneous Helmholz equation permit the same mapping between massive particle scattering and diffracted optical scalar waves. Scattering of narrow wave-packets finally recovers the classical trajectories. The analysis suggests that photothermal single particle microscopy infact measures photonic Rutherford scattering in specific limits.
\end{abstract}
\maketitle
\section{Introduction}
Almost exactly 100 years ago in the year 1911, Ernest Rutherford changed our picture of the atom by his famous theory on the scattering of positively charged $\alpha$-particles\cite{Rutherford1911}. In Rutherford scattering positively charged Helium nuclei are deflected by a Coulomb potential originating from positive nuclei of gold atoms as originally shown by Rutherford, Geiger and Mardsen\cite{GeigerMarsden1909}. This work has been a milestone in the discovery of the structure of the atom, revealing that most of the mass of an atom is concentrated in a tiny nucleus. Thus, Rutherford scattering is considered in each atomic physics lecture, treated in a classical framework to provide the characteristic angular distribution of scattered $\alpha$ particles. While a classical showpiece illustrating Rutherford scattering may be obtained from a paraboloidal hard wall-potential \cite{Wicher1965}, a direct display of the continuous trajectory or measuring a single deflection instead of the total cross-section remains difficult.
\begin{figure}[b]
\begin{center}\includegraphics [width=1.0\columnwidth]{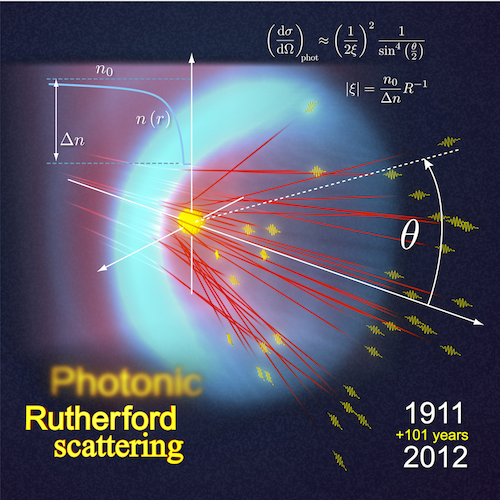} 
  \end{center}
\end{figure}
The classical theoretical predictions by E.\ Rutherford were later revisited to account for the detailed structure of the atom. While small impact parameters could be used to systematically probe the core-potential \cite{Temmer1989}, large impact parameters needed to additionally account for the electronic shielding \cite{Mantri1977} of the core Coulomb potential. Somewhat unexpectedly\cite{Barton1983,Yafaev1997}, the intricate \cite{Baryshevskii2004} quantum-mechanical spin-less treatment of the Coulomb $1/r$-potential predicted for all energies the same scattering-cross section as the classical theory \cite{LandauQM}.
Here we present the photonic analog of Rutherford scattering. It is given in the geometrical optics approximation (GOA) by the deflection of rays (the classical limit) or, in wave optics, as the diffraction of waves by a $1/r$-refractive index profile. This profile is provided by a heat point-source in a homogeneous medium. Such a point source may be a light-absorbing nano-particle \cite{Bohren1998Book} embedded in some medium which are used in photothermal single particle microscopy \cite{SelmkeDiffraction,Selmke2012ACSNano}. Experimental demonstrations of the effect can be achieved (see Section \ref{SectionPhotRutherford}).

The paper is structures as follows: In Section \ref{SectionFermat} the ray-optics treatment of the $1/r$-refractive index profile is presented and in Section \ref{SectionFermatSolution} the analytical solution derived. In Section \ref{SectionPhotRutherford} analogies of the found ray-optics solution are explored with respect to the classical non-relativistic and relativistic Rutherford scattering problem without radiation reaction. In Section \ref{SectionWaveMechanics} the wave-mechanical pictures are explored. Here, the correspondence between QM Coulomb scattering and the scalar optical field in the $1/r$-profile inhomogeneous refractive index field is revealed. Thereafter the correspondences to the classical pictures are established. Both an optical Fresnel-diffraction and a QM wave-packet formalism are used to achieve the necessary departure from the plane-wave limit. Finally, the found solutions are applied to photothermal microscopy and compared to previous experiments.

\section{Classical Limit: Fermats' Principle\label{SectionFermat}}
Obtainable through a variational principle with fixed path end-points which unifies Maupertius' (mechanics) and Fermat's (optics) variational principle, the following differential equation suitable for massive particles and light may be obtained\cite{Evans1996GRTphotpart,Nandi2001MatterWavesGRT,Evans1996}:
\begin{equation}\label{eq:fermatGeneral}
\frac{\mathrm{d}^2\mathbf{r}}{\mathrm{d}s^2}=\mathbf{\nabla}\left(\frac{1}{2}n^4\left(\mathbf{r}\right) v^2\left(\mathbf{r}\right)\right), \quad\quad \left| \frac{\mathrm{d}\mathbf{r}}{\mathrm{d}s}\right|=n\left(\mathbf{r}\right)^2 v\left(\mathbf{r}\right),
\end{equation}
with ${\bf r}$ being a vector on and $s$ a stepping parameter along the path. 
The difference in treating light or massive particles consists in the proper choice of the velocity $v\left(\mathbf{r}\right)$. In the latter case one may take $n=1$, such that Eq.\ (\ref{eq:fermatGeneral}) reduces to Newton's first law, Eq.\ (\ref{eq:Newton}), and thus also classical dynamics with the choice of the stepping parameter $\mathrm{d}s=\mathrm{d}t$, by setting $v^2/2=E/m - V/m$, i.e.\ the specific difference of total and potential energy per unit mass\cite{Evans1996}. Eq.\ (\ref{eq:fermatGeneral}) may even be used to describe relativistic gravitational mechanics in a static space-time metric by its corresponding non-unit refractive index\cite{Evans1996GRTphotpart,Evans1996}. To describe the paths of rays of light, Eq.\ (\ref{eq:fermatGeneral}) is to be supplemented by setting $v=c/n$, where $c$ is the vacuum speed of light. This case will correspond to Fermat's principle of the least optical path and allows the calculation of light trajectories through a spatially inhomogeneous refractive index field $n\left(\mathbf{r}\right)$. This picture provides a classical particle picture of light propagation and corresponds to the zero-wavelength limit of wave-optics\cite{BornWolf1980Book}. The result, Eq.\ (\ref{eq:fermat}), is the "F=ma"-optics developed by Evans et al. and explored by many others\cite{Evans1986,Evans1990,Rangwala2001,Drosdoff2005}
\begin{align}
{\rm mechanics:}\quad& m\frac{\mathrm{d}^2\mathbf{r}}{\mathrm{d}t^2}=-\mathbf{\nabla}V\left(\mathbf{r}\right), & \left| \frac{\mathrm{d}\mathbf{r}}{\mathrm{d}t}\right|=v\left(\mathbf{r}\right)\label{eq:Newton},\\
{\rm optics:}\quad& \frac{\mathrm{d}^2\mathbf{r}}{\mathrm{d}s^2}=\mathbf{\nabla}\left(\frac{1}{2}n^2\left(\mathbf{r}\right)\right), & \left| \frac{\mathrm{d}\mathbf{r}}{\mathrm{d}s}\right|=n\left(\mathbf{r}\right).\label{eq:fermat}
\end{align}
While the solution of positive energies to the Newton's equation of motion, Eq.\ (\ref{eq:Newton}), on a $1/r$-potential is known as Rutherford scattering (see Section \ref{SectionPhotRutherford}), we will now seek the physically achievable analogon in the optical domain. Consider a heat source that generates a temperature profile $T\left(r\right)=T_0+\Delta T\left(r\right)$ with
\begin{equation}
\Delta T\left(r\right)=P_{\rm abs}/\left(4\pi \kappa r\right),
\end{equation}
which, according to Fouriers law, decays with the inverse distance $r$ from the object to $T_0$ at infinite distance ($P_{\rm abs}$ and $\kappa$ are the absorbed power and the medium heat conductivity, respectively). This temperature profile results in the linear regime in the refractive index profile Eq.\ (\ref{eqn:RefracGradient}) that takes up the inverse distance dependence with the thermo-refractive coefficient $\mathrm{d}n/\mathrm{d}T$ as a proportionality factor,
\begin{equation}
n\left(\mathbf{r}\right) =n_0 + \frac{\mathrm{d}n}{\mathrm{d}T} \Delta T\left(\mathbf{r}\right) = n_0+\Delta n\frac{R}{r}\label{eqn:RefracGradient},
\end{equation}
where $n_0=n\left(T_0\right)$ is the unperturbed real-valued refractive index, $R$ the radius of the heat-source and $\Delta n=\Delta T\left(R\right)\mathrm{d}n/\mathrm{d}T$ a real-valued refractive index contrast. This is valid as long as the thermal conductivity of the finite-size heat source is larger than the mediums' conductivity. As we will demonstrate, the problem of finding the ray-trajectories fulfilling Eq.\ (\ref{eq:fermat}) is equivalent to the scattering by an unshielded Coulomb potential, i.e.\ Rutherford scattering.
A similar but rather artificial type of refractive index field, $n^2\!\left(r\right)={\rm const.}+2k/r$, has been shown to yield all types of Kepler-orbits for light in that medium\cite{Rangwala2001,Evans1996}. Also, effective refractive indices have been shown to mimic the path of light in gravitational fields as predicted by Einsteins theory of general relativity\cite{Felice1971,Deguchi1986,Nandi1995,Rangwala2001,Evans1996,Evans1996GRTphotpart,Nandi2001MatterWavesGRT}. For the weak gravitational field limit of the Schwarzschild metric $n\left(r\right)=1+2GMc^{-2}r^{-1}$ describes the null geodesics of light.

\section{Exact Solution\label{SectionFermatSolution}}
\begin{figure}[b]
\begin{center}\includegraphics [width=1.0\columnwidth]{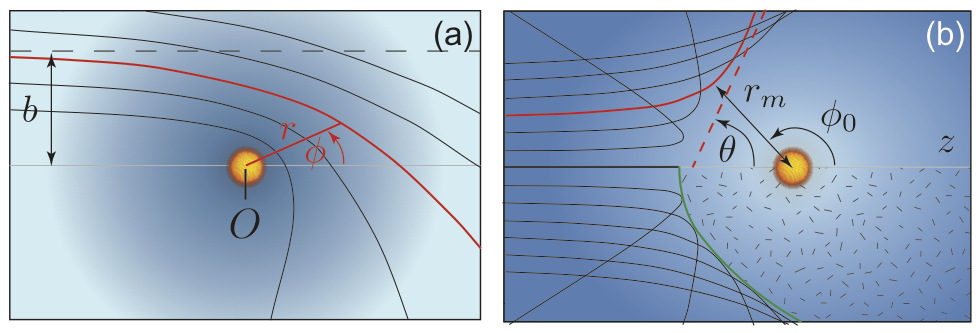} 
  \end{center}
  \caption{Annotated sketch of an exemplary ray trajectory (red) $r\left(\phi\right)$, Eq.\ (\ref{eq:path}), through the refractive index fields $n\left(r\right)$ with (a) $\Delta n >0$ and (b) $\Delta n < 0$ in Eq.\ (\ref{eqn:RefracGradient}).\label{fig:scat}}
\end{figure}
\begin{figure}[ht]
\begin{center}\includegraphics [width=1.0\columnwidth]{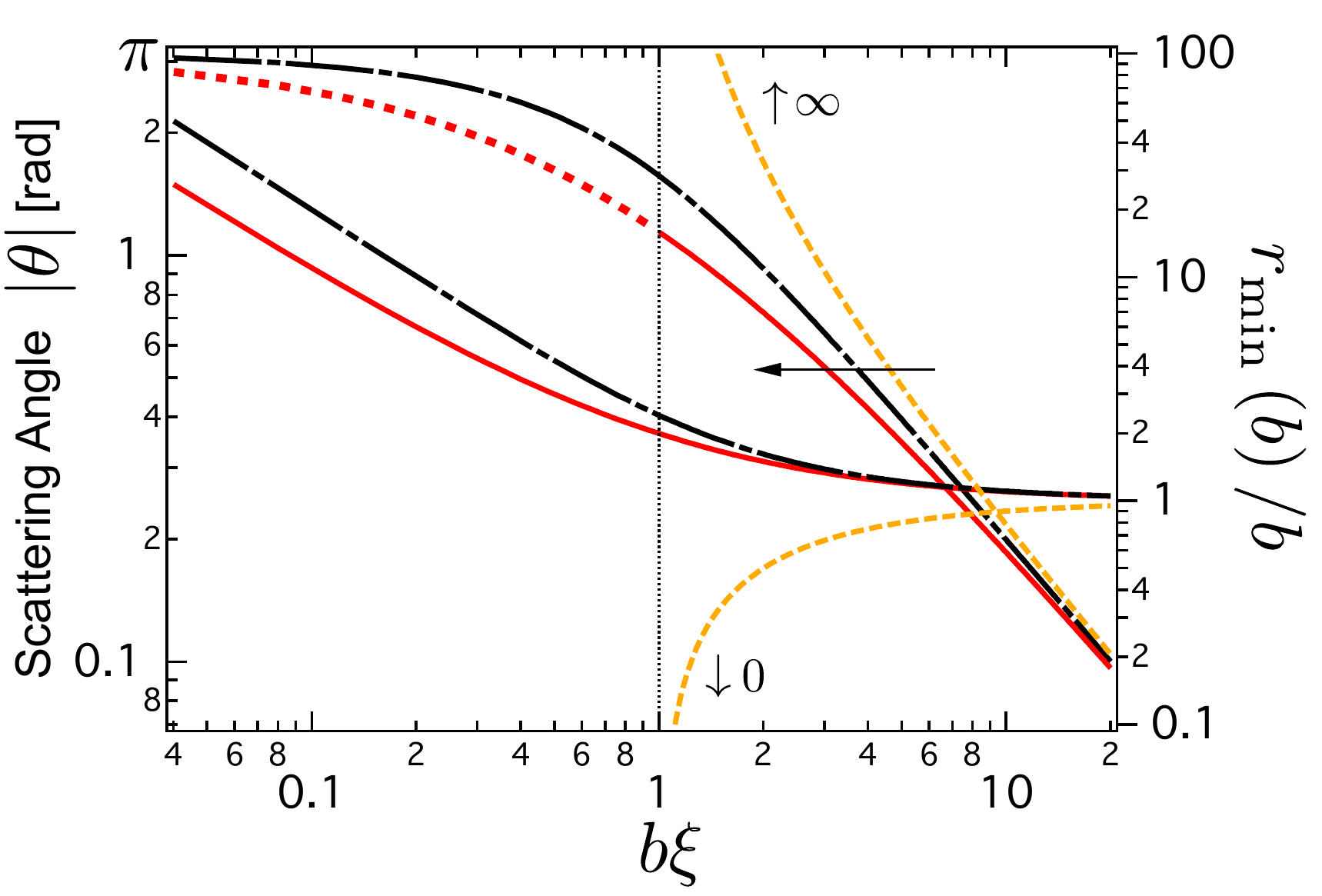}
  \end{center}
\caption{Absolute scattering/deflection angle $|\theta|$ (left axis) and the normalized distance of closest approach $r_{\rm min}\left(b\right)/b$ vs.\ impact parameter $b$ for fixed interaction strength $\xi^{-1}$. Black dashed-solid lines: Rutherford scattering, red lines: exact solution (orange dashed: attractive). Clearly visible is the effect of the additional attractive perturbative force allowing closer approaches and weaker deflections for the repulsive case ($\xi>0$) and stronger deflection in the attractive case ($\xi<0$). For large $b\xi\gg 1$ both results converge.\label{fig:ThetaRF}}
\end{figure}

Since, by symmetry, the trajectories will be confined to a plane (see Fig.\ \ref{fig:scat}), we use cylindrical coordinates $(r,\phi)$ where the acceleration takes the form $\mathbf{r}'' = \mathbf{\hat{r}}\left(r''-r\phi'^2\right) + \mathbf{\hat{\theta}}\left(r\phi''+2r'\phi'\right)$ and the gradient reads $\mathbf{\nabla}n = \mathbf{\hat{r}}\,\partial_r n +\mathbf{\hat{\theta}}\,r^{-1}\partial_\theta n=n^{-1}\mathbf{\nabla} n^2/2$. The prime denotes differentiation with respect to the stepping parameter $s$. Fermats' least optical path principle Eq.\ (\ref{eq:fermat}) then gives two equations, Eq.\ (\ref{eq:rhat}) for the radial coordinate and Eq.\ (\ref{eq:thetahat}) for the angular coordinate:
\begin{eqnarray}
\mathbf{\hat{r}}: \qquad & r'' - r\phi'^2 &= \underset{\rm attractive / repulsive}{-n_0\Delta n R\frac{1}{r^2}} \,\,\, \underset{\rm attractive}{\underline{-\Delta n^2 R^2\frac{1}{r^3}}}\label{eq:rhat}\\
\mathbf{\hat{\theta}}:\qquad & r\phi''+2r'\phi' &= 0\label{eq:thetahat}
\end{eqnarray}
The above set of coupled differential equations is equivalent to the perturbed Kepler problem with its precessing orbit solutions \cite{Sivardiere1986}.  Equation (\ref{eq:thetahat}) yields the conserved optical angular momentum $L_z=r^2\phi'$, i.e.\ $L_z'=0$. Now, the formula of Bouguer\cite{Evans1986} allows to express this quantity at any point along the trajectory as $L_z=r\sin\left(\phi\right)|\mathrm{d}\mathbf{r}/\mathrm{d}s|$, such that with Eq.\ (\ref{eq:fermat}) we find at infinite distance $L_z=b n_0$. The parameter $b>0$, the so called impact parameter, is the distance of the approaching parallel ray to the optical axis (see Fig.\ \ref{fig:scat}a). The differential Eq.\ (\ref{eq:rhat}) is the analogue to the mechanical radial force equation and shows an inverse radius squared interaction, which is either attractive or repulsive depending on the sign of $\Delta n$, and a perturbation by a inverse radius cubed term (underlined in the following). To solve it for $r\left(\phi\right)$, a change of differentials is needed. Applying $\mathrm{d}/\mathrm{d}s=\phi'\mathrm{d}/\mathrm{d}\phi=L_z r^{-2}\mathrm{d}/\mathrm{d}\phi$ twice, and introducing the inverse radius variable $u=1/r$ one finds the following relation
\begin{equation}
r''=\frac{\mathrm{d}^2 u^{-1}}{\mathrm{d}s^2}=L_z^2u^2\frac{\mathrm{d}}{\mathrm{d}\phi}\left(u^2\frac{\mathrm{d}}{\mathrm{d}\phi}
\frac{1}{u}\right)
=-L_z^2 u^2 \frac{\mathrm{d}^2u}{\mathrm{d}\phi^2},
\end{equation}
which transforms Eq.\ (\ref{eq:rhat}) into
\begin{equation}
-L_z^2 u^2 \frac{\mathrm{d}^2u}{\mathrm{d}\phi^2} - L_z^2 u^3=-n_0\Delta n R u^2\,\underline{-\Delta n^2 R^2u^3} \label{eq:u2}
\end{equation}
We now introduce the variable
\begin{equation}
\xi=-\frac{n_0}{\Delta n R},
\end{equation}
which is a measure for the inverse strength of the heat induced refractive index gradient and encodes the polarity of the interaction in such a way that a positive sign of $\xi$ corresponds to repulsion. Equation (\ref{eq:u2}) then becomes, after rearranging and collecting of the terms linear in $u$,
\begin{equation}
\frac{\mathrm{d}^2u}{\mathrm{d}\phi^2} +u \left(1\,\underline{-\,b^{-2}\xi^{-2}}\right)= -\xi^{-1}b^{-2}.\label{eq:u}
\end{equation}
If the refractive index in the medium is homogeneous, i.e.\ $\xi=\infty$, the harmonic oscillator differential equation with unit angular frequency emerges and the correct solution fulfilling the boundary conditions is $u=r^{-1}=b^{-1}\sin\left(\phi\right)$. In cartesian coordinates $y=r\sin\left(\phi\right)$ this is a straight line parallel to the optical axis at a distance $b$, which of course is the unperturbed ray, see dashed line in Fig.\ \ref{fig:scat}a. If the perturbation is nonzero, and requiring for the moment that $|b\xi|>1$, Eq.\ (\ref{eq:u}) has the form of the familiar harmonic oscillator differential equation plus a constant, $u''+c_1 u = -c_2$ with positive $c_1$. It is solved by $u=\frac{c_2}{c_1}\left[e\cos\left(\sqrt{c_1} \left(\phi-\phi_0\right)\right)-1\right]$ with the yet to be determined constants $e$ and $\phi_0$. Equation (\ref{eq:u}) is therefore solved by
\begin{equation}\label{eq:path}
r\left(\phi\right)=\frac{p}{e \cos\left(\gamma \left[\phi-\phi_0\right]\right)-1},
\end{equation}
where eccentricity is allowed to be either positive or negative, and with the parameters
\begin{equation}
\left. \begin{array}{ll}
	p &= \left[b^2\xi^2-1\right]/\xi\\
	\gamma^2 &= 1-b^{-2}\xi^{-2}\\
	\end{array}\right\}.\label{eq:Parms1}
\end{equation}
Mathematically, the orbits described by Eq.\ (\ref{eq:path}) represent perturbed hyperbolic trajectories with the particle being the exterior ($\xi>0$) or interior ($\xi<0$) focus\cite{Rangwala2001}, see Fig.\ \ref{fig:scat}a,b. More exactly, they are epispirals, a special case of so-called Cotes's spirals. Such orbits may show peculiar behavior, such as multiply revolving trajectories for $\xi<0$, when the perturbation-parameter $\gamma$ approaches zero (see Fig.\ \ref{fig:ThetaRF}) and were already discussed by the grandson of Charles Darvin, C. G. Darwin, in the context of relativistic Rutherford scattering of electrons in 1913\cite{Darwin1913}. Also, somewhat later in 1916, Sommerfeld in his relativistic corrections to the Hydrogen spectra encountered the bound form of such orbits for the electron\cite{Sommerfeld1996relOrbit,LandauFeldtheorie,Boyer2004,Gliwa1996}. To obtain the eccentricity $e$ we reconsider the particular choice of the stepping parameter in Eq.\ (\ref{eq:fermat}), and write again in cylindrical coordinates:
\begin{equation}
\left| \mathbf{r}'\right|=n \quad \rightarrow \quad r'^2+r^2\phi'^2=n\left(r\right)^2.\label{eq:rmin}
\end{equation}
The radius of closest approach is obtained by setting $r'=0$, and yields $r_{\rm m}=b+\xi^{-1}$. Again, angular momentum conservation $\phi'=L_z r^{-2}$ was used. Comparison of this expression to the corresponding minimum radius as described by Eq.\ (\ref{eq:path}), $r_{\rm m}=p/\left(e-1\right)$ at the angle of closest approach $\phi=\phi_0$, yields the eccentricity $e=b\xi$. Setting the denominator of Eq.\ (\ref{eq:path}) to zero yields the extreme angles $\theta_{\infty}^{\pm}=\pm|\gamma^{-1}|\arccos\left(1/e\right)+\phi_0$. Requiring that the ray approaches parallel to the optical axis from negative infinity, see Fig.\ \ref{fig:scat}b, i.e.\ $\theta_{\infty}^{+}=\pi$, will orient the solution Eq.\ (\ref{eq:path}) according to the imposed initial conditions. We then find the angle of closest approach:
\begin{equation}
\left. \begin{array}{ll} 
	e &= b\xi\\
	\phi_0 &= \pi - |\gamma^{-1}|\arccos\left(e^{-1}\right)
\end{array}\right\}\label{eq:Parms2}
\end{equation}
The parameters in Eqs. (\ref{eq:Parms1}) and (\ref{eq:Parms2}) together with Eq.\ (\ref{eq:path}) now fully determine the ray-trajectory. The scattering angle $\theta=\theta_{\infty}^{-}$, i.e. the deflection angle of an incoming horizontal ray, may be expressed as $\theta=2\phi_0-\pi$. We finally note that the differential Eq.\ (\ref{eq:u}) can also be obtained from Binet's orbit equation\cite{Sivardiere1986,Eliseo2006} with the correct identification of the force terms as given by Eq.\ (\ref{eq:fermat}).

The previous treatment relied on the assumption, which is however valid in practical situations, that $|b\xi| > 1$. If the impact parameter gets very small, $\gamma$ would become imaginary. This situation is solemnly due to the presence of the attractive inverse cubic interaction term which dominates the inverse squared term at small distances, see Eq.\ (\ref{eq:rhat}). Instead of Eq.\ (\ref{eq:u}), we must then solve the following differential Eq.\ :
\begin{equation}
\frac{\mathrm{d}^2u}{\mathrm{d}\phi^2} - u \left(b^{-2}\xi^{-2} - 1\right)= -\xi^{-1}b^{-2}.\label{eq:usmall}
\end{equation}
It has the form $u'' - c_1 u = -c_2$ with positive $c_1$ and is solved by $u=\frac{c_2}{c_1}\left[e\cosh\left(\sqrt{c_1} \left(\phi-\phi_0\right)\right) + 1\right]$, where we have chosen the hyperbolic cosine for now and will consider the general solution hereafter. Therefore,
\begin{equation}\label{eq:pathSmall}
r_{r}\left(\phi\right)=\frac{p_r}{e_r \cosh\left(\gamma_{r} \left[\phi-\phi_{0,r}\right]\right) + 1},
\end{equation}
with the positive perturbation parameter $\gamma_r > 1$ determined by $\gamma_{r}^2 = b^{-2}\xi^{-2} - 1 = -\gamma^2$ and $p_r=\left[1-b^2\xi^2\right]/\xi=-p$. The only admittable solution for an approach from infinity is for an eccentricity to be within $-1 < e_r < 0$.
In this situation $r_{m,r} = b+\xi^{-1}>0$ for $\xi>0$ (and only for the \textbf{repulsive} case) is achieved at $\phi=\phi_0$ and indeed yields $e_r = -b\xi$ in the desired range. Setting the denominator of Eq.\ (\ref{eq:pathSmall}) to zero one finds the extreme angles $\theta_{\infty,r}^{\pm}=\pm |\gamma_r^{-1}|{\rm arccosh}\left(b^{-1}\xi^{-1}\right) + \phi_0$ such that again we correctly orient the solution with the choice $\theta_{\infty,r}^{+}=\pi$ and thereby $\phi_{0,r} = \pi - |\gamma_r^{-1}|{\rm arccosh}\left(b^{-1}\xi^{-1}\right)$. Here, too, the deflection angle is $\theta_r=2\phi_0-\pi \approx \pi + 2b\xi\ln\left(b\xi/2\right) + \mathcal{O}\left(b^2\xi^2\right)$ and its limit is $\theta\rightarrow \pi$ as $b\xi \rightarrow 0$, which corresponds to a perfect retroreflection for a head-on impact of a ray onto the lens. Both $\left\{\phi_{0,r},\theta_r\right\}$ are smooth continuations of $\left\{\phi_{0},\theta\right\}$ found earlier. Infact, allowing the cosine to have a complex argument with $\gamma_r=i\gamma$ in Eq.\ (\ref{eq:path}), the same solution is obtained.

We now seek the general solution of Eq.\ (\ref{eq:usmall}):
\begin{equation}\label{eq:pathSmallGeneral}
r_{a}\left(\phi\right)=\frac{p_a}{e_{a,1} \exp\left(\gamma_{a}\phi\right) + e_{a,2} \exp\left(-\gamma_{a}\phi\right) + 1},
\end{equation}
This ansatz now allows the incoming ray to have the correct distance at infinity, e.g.\ $\lim_{\phi \to \pi} \sin\left(\phi\right) r_{s}\left(\phi\right) = b$, and gives the set of two two generalized eccentricities:
\begin{equation}
\left. \begin{array}{ll}
	e_{a,1} &= -e^{-\pi\gamma_a} \left[\frac{p_a+b\gamma_a}{2b\gamma_a}\right],\\
	e_{a,2} &= -e^{\pi\gamma_a} \left[1-\frac{p_a+b\gamma_a}{2b\gamma_a}\right].
\end{array}\right\}\label{eq:Parms2}
\end{equation}
Solution (\ref{eq:pathSmallGeneral}) works for both the \textbf{attractive} case, hence the subscript ${}_a$, and the repulsive case. In the former case the solution is a true mixture of the hyperbolic sine and cosine which describes trajectories approaching from infinity and falling within a finite time into the coordinate origin. It does so without a closest distance $r_m$ and coming from the $b\xi<-1$ case the rays revolve evermore vigorously around the origin. Both phenomena continue the limiting behavior of Eq.\ (\ref{eq:path}) where the closest approach distance goes to zero and the scattering angle $\theta$ diverges to infinity as $|b\xi|\rightarrow 1$, see Fig.\ (\ref{fig:ThetaRF}). In the case of repulsive interaction the solution given above guises the solution involving only the hyperbolic cosine found earlier, i.e.\ Eq.\ (\ref{eq:pathSmall}).

Similar to certain cases of relativistic point-particle Kepler mechanics\cite{Boyer2004}, these two special solutions involving the hypergeometric functions corresponds to special types of Cotes spirals.


\section{Photonic Rutherford Scattering\label{SectionPhotRutherford}}

As the refractive index change itself is typically small for most materials ($|\Delta n|\approx 10^{-3}$), and since $b> R$ for the incoming rays, the product $| b\xi |\gg 1$ is a large number. This allows allows us to approximate Eq.\ (\ref{eq:path}) by
\begin{equation}\label{eq:Rutherford_Photon}
r\left(\phi\right)\approx \frac{|\xi| b^2}{\sqrt{b^2\xi^2+1}\cos\left(\phi-\phi_0\right) \pm 1},
\end{equation}
where $\pm$ is the sign opposite of $\xi$, which now shows complete equivalence to the classical (non-relativistic) Rutherford scattering solution of Eq.\ (\ref{eq:Newton}) on the potential $V\left(r\right)=C r^{-1}$ (attractive: $C<0$, repulsive: $C>0$),
\begin{equation}\label{eq:Rutherford_Coulomb}
 r_{\rm RF}\left(\phi\right)=\frac{2Eb^2/ |C|}{e\cos\left(\phi-\phi_0\right) \pm 1}
 \end{equation}
where the notion is such that attractive interaction is represented by the upper and repulsive interaction by the lower sign, respectively. The scattering parameters are
\begin{equation}
\left. \begin{array}{ll} 
	E&= m v_0^2/2\\
	C&=q_1 q_2/\left(4\pi\epsilon_0\right)\\
	e^2&=4E^2 b^2 C^{-2}+1, \,\, e\ge 0\\
	\phi_0&=\pi \pm \arccos\left(1/e\right)
\end{array}\right\},\label{eq:ParmsRutherford}
\end{equation}
and describe the total energy $E$ and mass $m$ of the scattered particle, and $e^2$ the squared eccentricity of the orbit, $C$ the Coulomb force constant for the two charges $q_{1,2}$ that describes the mechanical force $\mathbf{F}\left(r\right)=-C r^{-2}\mathbf{\hat{r}}$. The scatterer is assumed to be fixed here, i.e.\ has an infinite mass as compared to the scattered particle. The angular momentum of the particle relative to the scatterer at the origin is $L=m v_0 b$, while its specific angular momentum (twice the areal velocity) is $L_z=L/m$.
The deflection of photons by a weak gradient index lens generated by a heated point-like absorber, described by Eq.\ (\ref{eq:Rutherford_Photon}), is thus the complete photonic analogon of Rutherford scattering of $\alpha$ particles on a single nucleus, Eq.\ (\ref{eq:Rutherford_Coulomb}). $V \rightarrow -n\left(r\right)^2\!\!/2 + n_0^2/2\approx n_0^2\, \xi^{-1} r^{-1}$ can therefore be identified as the photonic analogon of the potential energy decaying to zero at infinite distance, $E \rightarrow n_0^2/2$ being the total energy and $C\rightarrow -n_0 \Delta n R$ is the equivalent of the Coulomb force constant, as can be inferred from Eq.\ (\ref{eq:rhat}). The form of Eq.\ (\ref{eq:fermat}) also requires the mass to be set to unity $m=1$ in optics. Hence, all further equations, e.g. the differential scattering cross section $\left(\frac{\mathrm{d}\sigma}{\mathrm{d}\Omega}\right)$ unravelling the famous $\sin^{-4}(\theta/2)$ dependence, or the total cross-section $\sigma_{>\Theta}$ of scattering by an angle larger than some angle $\Theta$ can be obtained using these equivalences and the substitution $2E/C \rightarrow \xi$. 

The observation that the total energy is positive requires a few comments. Typically\cite{Evans1986,Evans1996} it is stated that Eq.\ (\ref{eq:fermat}), $\frac{1}{2}|\mathbf{r}'|^2-\frac{1}{2}n^2=0$, corresponds to the equation for the total energy analogon, comprised of a kinetic energy term $\frac{1}{2}|\mathbf{r}'|^2$ and a potential energy term $-\frac{1}{2}n^2$, and thus the total energy in the optical case is equivalent to the mechanical scenario at zero energy $E=0$. However, due to the inclusion of the additional constant shift ($+n_0^2/2$) of the potential energy scale in $V$, necessitated by including $n_0$ in Eq.\ (\ref{eqn:RefracGradient}), we find that here the mechanical zero-energy scenario does not represent the optical problem at hand (cf.\ footnote 15 of the Evans et al. "F=ma"-optics paper \cite{Evans1986}). Indeed, the parabolic unit-eccentricity orbits of zero-energy scattering is not the found (approximate) solution for the ray-trajectory. Here, $\frac{1}{2}|\mathbf{r}'|^2 + V - n_0^2/2 = 0$ and $E$ can be identified with the first two terms yielding $E=n_0^2/2$ to be taken as the mechanical total energy analogon. Thus, only the unbound (hyperbolic) trajectories from classical mechanics are attainable for $n_0\ne 0$. 

The discrepancy by a factor of $2$ between the expression for the distance of closest approach $r_C$ and the exact value from Eq.\ (\ref{eq:rmin}), $r_m\left(0\right)=\xi^{-1}$, stems from the fact that for $b\rightarrow 0$ the validity of the approximation $b\xi \gg 1$ and thus Eq.\ (\ref{eq:Rutherford_Photon}) breaks down. For a repulsive potential Eq.\ (\ref{eq:pathSmall}) then passes the point of closest approach. The same argument explains the difference between $r_{\rm min}\left(b\right)$ and $r_m\left(b\right)$. For the repulsive case ($\Delta n<0$), Eq.\ (\ref{eq:rhat}) shows an additional attractive inverse radius-cubed interaction resulting in a closer approach. Solving Eq.\ (\ref{eq:rmin}) without such a term yields the photonic $r_{\rm min}\left(b\right)$-value listed in Table \ref{table:Parallels}. The exact trajectories will thus penetrate the classical Rutherford shadow region given by the paraboloid $r_s=4\xi^{-1}/\left[1-\cos\left(\phi\right)\right]$. For large $b\xi\gg 1$ the two expressions coincide.

\renewcommand{\arraystretch}{2}
\renewcommand{\tabcolsep}{9pt}

\begin{table}[ht]
\centering
\begin{tabular}{|c|c|c|}
\hline
quantity & photonic scattering & Coulomb scattering \\
\hline
$v\left(\mathbf{r}\right)$ & $n\left(\mathbf{r}\right)$ & $v\left(\mathbf{r}\right)$\\
$V\left(r\right)$ & $n_0^2 \xi^{-1} r^{-1}$ & $C r^{-1}$\\
$L$ & $n_0 b$ & $m v_0 b$\\
$C$ & $-n_0 R \Delta n$ & $q_1 q_2/\left[4\pi \epsilon_0\right]$\\
$E$ & $n_0^2/2$ & $mv_0^2/2$\\
$r_C$ & $2\xi^{-1}$ & $C/E$\\
$r_{\rm min}\left(b\right)$ & $\frac{1}{\xi} + \frac{1}{\xi}\sqrt{b^2\xi^2+1}$ & $\frac{r_C}{2} + \frac{r_C}{2}\sqrt{4b^2r_C^{-2}+1}$\\
$r_s\left(\phi\right)$ & $\frac{4\xi^{-1}}{1-\cos\left(\phi\right)}$ & $\frac{2 C/E}{1-\cos\left(\phi\right)}$\\
$\cot\left(\frac{\theta}{2}\right)$ & $b\xi$ & $2Eb/C$\\
$\sigma_{>\Theta}$ & $\frac{\pi}{\xi^2}\left[\frac{1+\cos\left(\Theta\right)}{1-\cos\left(\Theta\right)}\right]$ & $\pi\left(\frac{C}{2E}\right)^2\left[\frac{1+\cos\left(\Theta\right)}{1-\cos\left(\Theta\right)}\right]$\\
$\left(\frac{\mathrm{d}\sigma}{\mathrm{d}\Omega}\right)$ & $\left(\frac{1}{2\xi}\right)^2 \sin^{-4}\left(\frac{\theta}{2}\right)$ & $\left(\frac{C}{4E}\right)^2\sin^{-4}\left(\frac{\theta}{2}\right)$\\
\hline
\end{tabular}
\vspace{0.2cm}
\caption{Correspondence table showcasing different expressions in photonic and Rutherford/Coulomb scattering.}
\label{table:Parallels}
\end{table}

The before-mentioned similarity to relativistic motion in a $1/r$-potential\cite{Darwin1913,Boyer2004,Gliwa1996} (cf.\ paragraph \S 39 of ref \cite{LandauFeldtheorie}), which was used by Arnold Sommerfeld to give the fine-splitting of the Hydrogen line-spectrum\cite{Sommerfeld1996relOrbit}, may be brought to a correspondence with the photonic problem here using the previous substitutions complemented by the additional rule $c\rightarrow n_0$ and with $E=mc^2 \rightarrow n_0^2$ now replacing the total energy including the rest-mass. However, in optics there is no such distinction between relativistic and non-relativistic treatments of light, such that it shall not be implied that optics corresponds to mechanics in its special relativistic form. Still, the analogy is between non-relativistic mechanics and optics as embodied in Eq.\ (\ref{eq:fermat}). Furthermore, these relativistic Rutherford scattering solutions should not be taken as necessarily being more accurate since also here radiation reaction as embodied in the Lorentz-Abraham-Dirac equation are not considered (see Huschilt et al. \cite{Huschilt1976,Huschilt1978} or Aguiar et al. \cite{Aguiar2009} and references therein).

\begin{figure}[ht]
\begin{center}\includegraphics [width=1.0\columnwidth]{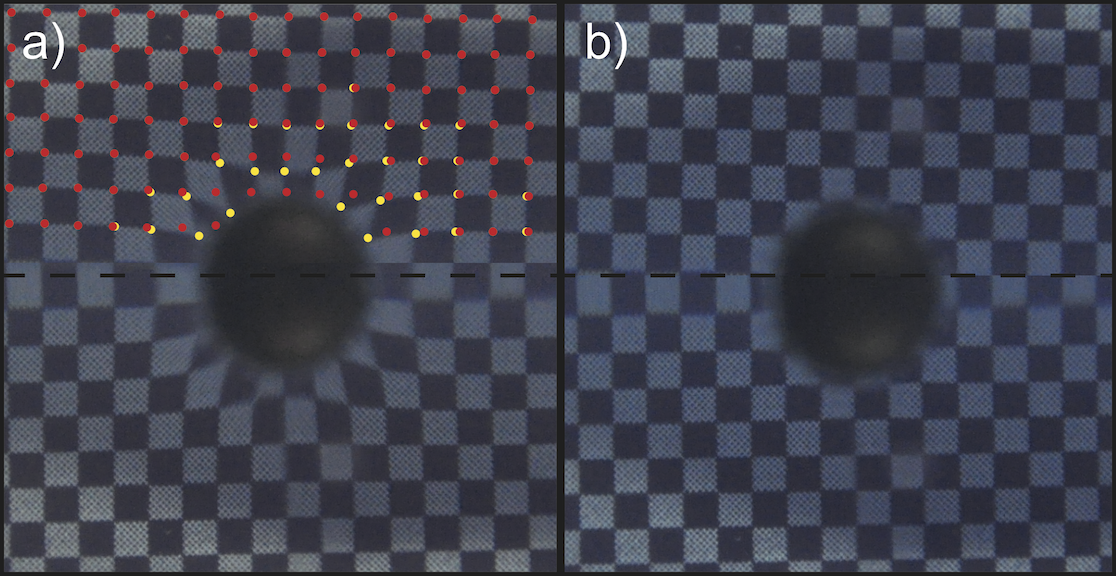}
  \end{center}
\caption{Visual effect of the photonic scatterer (laser-heated metal sphere in transparent resin). An image photographed through the medium is warped. The deflection of rays gives the illusion of a crunching of the image. Warped image a) and the initial image b) were mirrored along the horizontal due to the layering turbidity from the manufacturing process.\label{fig:LensEffect}}
\end{figure}

\begin{figure}[h!]
\begin{center}\includegraphics [width=1.0\columnwidth]{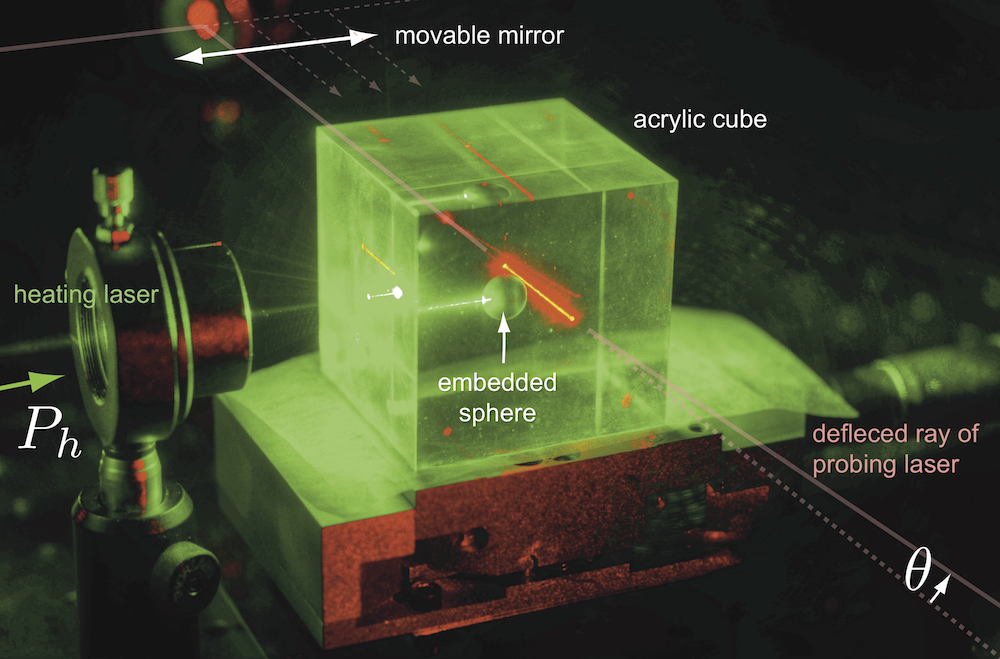}
  \end{center}
\caption{Macroscopic Experiment on single Rutherford-like photonic scatterer (black-body sphere with a small hole). For a typical polymer medium, $\kappa\approx 10^{-1}{\rm W m}^{-1}{\rm K}^{-1}$. Experimental conditions of $P=1{\rm W}$, $R=0.5{\rm mm}$ give a temperature of $\Delta T_0\approx 100{\rm K}$. With $\mathrm{d}n/\mathrm{d}T\approx3\times 10^{-4}$, $n_0=1.5$ a deflection angle of $\theta\approx 3^{\circ}$ is expected. When $D$ is chosen large enough, one may easily observe the deflection $\Delta x=D\tan\left(\theta\right)$ on a screen.\label{fig:MacroExp}}
\end{figure}

Rutherford scattering is generally considered for multiple scattering on many nuclei with random impact parameters. Thus, the measurable cross-section delivers the same results for attractive or repulsive Coulomb interactions. While the latter results are obtained by a classical and QM wave-mechanics, similar prediction for diffraction on multiple refractive index profiles have been made\cite{Vigasin1977}. The photonic equivalent can, however, be tested easily on a single scattering center, allowing to access even the sign of the interaction, i.e.\ the sign of $\mathrm{d}n/\mathrm{d}T$, with the help of simple photodetector. For this, a macroscopic experiment with a metal sphere embedded in a transparent resin as shown in Fig.\ \ref{fig:MacroExp} can be setup. Upon heating of the central sphere by a high-power laser one can measure the deflection of paraxial thin laser-beams according to $\cot\left(\theta/2\right)=b\xi$.
Also, quite naturally the photonic Rutherford scattering can be seen by the unaided eye directly. Viewing an object through such a medium containing a heat point-source will cause the viewed object to appear warped according to the extrapolated path as seen in Fig.\ \ref{fig:LensEffect}. As noted before, a refractive index profile of $n\left(r\right)=1+2GMc^{-2}r^{-1}$ describes gravitational lensing\cite{Felice1971,Deguchi1986,Nandi1995,Rangwala2001,Evans1996,Evans1996GRTphotpart,Nandi2001MatterWavesGRT}. Therefore, the observed distortion nicely model for instance the famous Einstein ring phenomenon if a material with $\Delta n>0$ is used (some types of glasses such as N-PK51 have this property, c.f. the TIE19 data sheet "Temperature Coefficient of the Refractive Index" of the manufacturer \textit{Schott}). A computer program interactively visualizing these effects is publicly available on the authors' webpage.

Photothermal single particle microscopy also provide measurements on single photonic Rutherford scatterers\cite{Selmke2012ACSNano}. A simple formalism starting from the ray-optics results presented here have been used to provide a semi-quantitative minimal model for photothermal lensing microscopy of heatable metallic nanoparticles\cite{Selmke2012ABCD}.


\newpage

\section{Wave mechanical Rutherford scattering\label{SectionWaveMechanics}}
Similar to scattered alpha particles, also photons obey the wave-particle duality. While they interact only very weakly among each other, their interaction with matter is described in its strength by the dielectric function $\epsilon$. The dielectric function thus defines a "photonic potential" manipulating the propagation of photons or optical scalar fields in the simplest form of wave optics. Similarly, Quantum Mechanics is the wave-description of matter. While the equivalence of the wave-optics treatment of the Coulomb scattering problem with the classical has been shown for the plane-wave case, we will here demonstrate the correspondence also in optics and with finite beams (of either particles, or rays).
Concepts from scalar wave optics have been applied to Quantum problems ever since and show the close relation of both. As seen in nuclear scattering experiments\cite{Frahn1966,Frahn1971}, molecular interferrometry data \cite{Kurtsiefer1997}, or atomic aperture diffraction experiments \cite{Goussev2012}, interference effects for instance may conveniently be described by Fresnel diffraction. Further mappings have been found between paraxial wave-optics and the Schr\"odinger equation in two dimensions\cite{Marte1997,Marte1998,Steuernagel2005_QMosci}. We will now show the equivalence between our recent scalar wave optics treatment of the diffraction by the refractive index profile $n\left(r\right)$, Eq.\ (\ref{eqn:RefracGradient}), and the QM problem of scattering on a bare Coulomb potential.

\subsection{Plane wave scattering / diffraction\label{SubSectionPWDiffraction}}
The Schr\"odinger equation (SE) for scattering on a Coulomb potential $V\left(r\right)=C r^{-1}$ reads $-\frac{\hbar^2}{2m}\nabla^2\Psi_C + \frac{C}{r}\Psi_C=E\Psi_C=\frac{1}{2}m v_0^2\Psi_C$. The wave vector $k$ of the incident particle-wave is defined through the de Broglie relation $\hbar k=m v_0$. The time-independent SE may then be rewritten as
\begin{equation}
\nabla^2\Psi_C+\left[k^2-\frac{2 \nu k}{r}\right]\Psi_C=0,\label{SchroediEq}
\end{equation}
where the introduced interaction parameter $\nu=\frac{C k}{2E}$ denotes the strength and polarity of the potential. A positive value of $\nu>0$ corresponds to a repulsive, and a negative $\nu<0$ to an attractive potential. The analytical solution, first given by Gordon in 1928\cite{Gordon1928}, to the equation is achieved by the ansatz $\Psi_C\left(\mathbf{r}\right)=e^{i k z}f\left(r-z\right)$, wherein $r^2=\rho^2+z^2$. The complex-valued function $f$ describes the perturbation of the incoming plane wave. Inserting this ansatz and the Laplacian in cylindrical coordinates into the SE leads to
\begin{equation}
\left[\frac{\partial^2}{\partial \rho^2}+\frac{1}{\rho}\frac{\partial}{\partial \rho}+2ik\frac{\partial}{\partial z}+\frac{\partial}{\partial z^2}-\frac{2\nu k}{r}\right]f\left(r-z\right)=0.
\end{equation}
Defining the function $g\left(x\right)$ by $f\left(r-z\right)=g\left(x\right)$ with the variable substitution $x=ik\left(r-z\right)$ one obtains Eq.\ (\ref{eq:Hyp1f1}), which is the hypergeometric differential equation for $g\left(x\right)$
\begin{equation}
\left[x\,\frac{\mathrm{d}^2}{\mathrm{d}x^2}+\left(1-x\right)\frac{\mathrm{d}}{\mathrm{d}x}+i\nu\right]g\left(x\right)=0\label{eq:Hyp1f1}.
\end{equation}
Therefore, the solution wave function $\Psi_C$ for the case of an incident plane wave may be written down \cite{LandauQM}:
\begin{equation}
\Psi_C\left(\mathbf{r}\right)=e^{-\frac{\pi}{2}\nu}e^{ikz}\Gamma\left(1+i\nu\right){}_{1}F_{1}\left(-i\nu;1;ik\left(r-z\right)\right)\label{SchroediExactSolution}
\end{equation}
The pre-factors ensure a normalization to unity $|\Psi_C|^2=1$ at large distances $z\rightarrow \infty$ from the scatterer. Also, the wave-function reduces to the incoming plane wave for vanishing perturbation, i.e.\ $\Psi_C\left(\mathbf{r}\right)=e^{ikz}$ for $\nu=0$. An asymptotic expansion of the confluent hypergeometric function for large $k \eta=k\left(r-z\right)\rightarrow \infty$ allows the wave function to be separated into a scattered spherical wave with angle-dependent amplitude $f$ and a plane wave resembling the form $e^{ikz}+f\left(\theta\right)e^{ikr}/r$, although both terms will include logarithmic phase distortions due to the long-range character of the Coulomb potential\cite{LandauQM}. Apart from corrections vanishing for $r\rightarrow \infty$, the scattering cross-section reads
\begin{equation}
\frac{\mathrm{d}\sigma}{\mathrm{d}\Omega}=|f\left(\theta\right)|^2=\left(\frac{\nu}{2k}\right)^2\frac{1}{\sin^4\left(\theta/2\right)}
\end{equation}

\begin{figure}[ht]
\begin{center}\includegraphics [width=1.0\columnwidth]{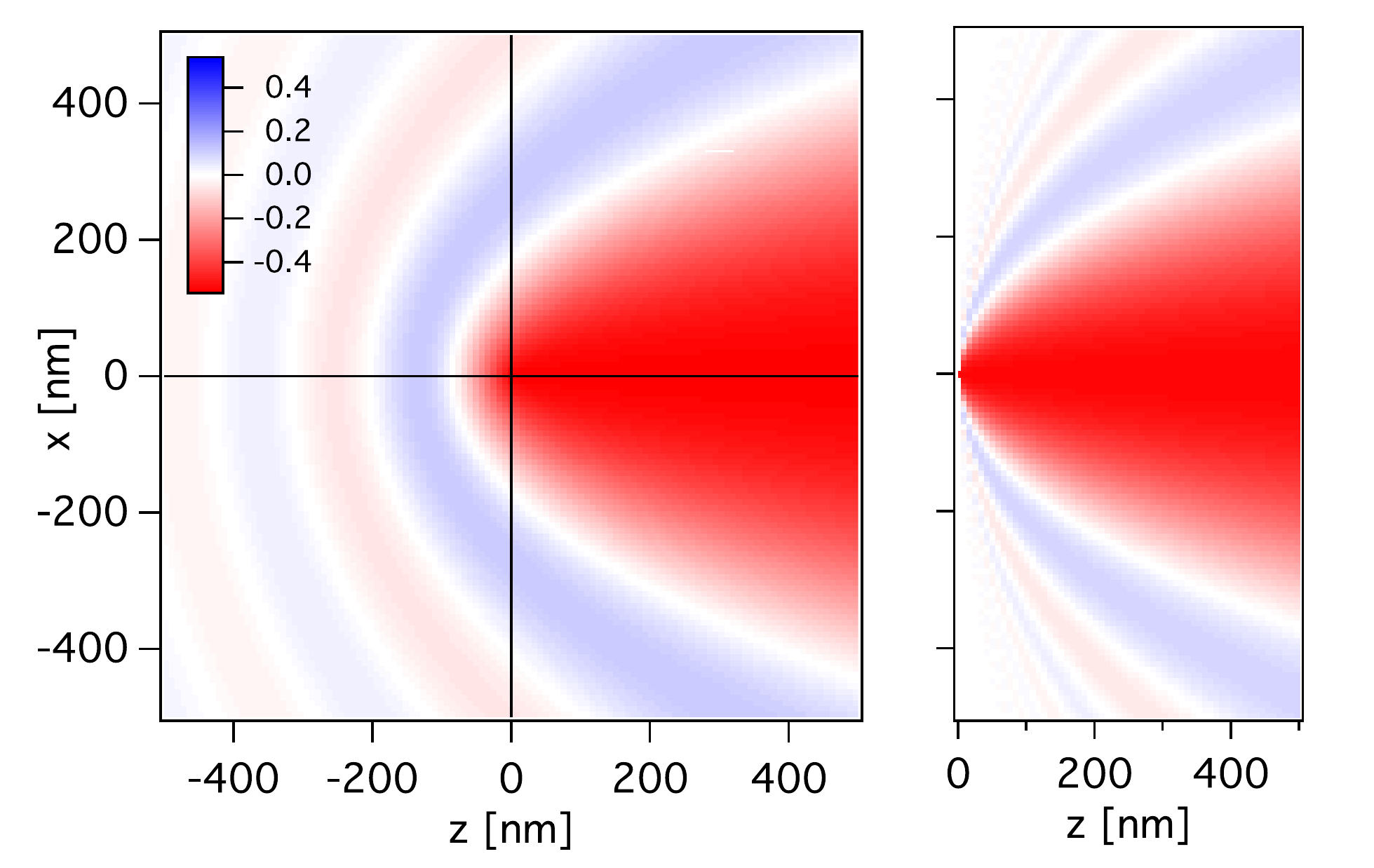}
  \end{center}
\caption{Top: $|\Psi_C\left(r,z\right)|^2-1$ from Eq.\ (\ref{SchroediExactSolution}) and the GOA shadow lines (green) and the $x\left(z\right)=z\tan\left(\theta_0\right)$ lines (blue) enclosing the "near" zone. Bottom left: Zoom. Bottom right: Eq.\ (\ref{EqnFresnelPW}). The first line-scan position of Fig.\ \ref{FigPsiScans} is indicated by the dashed line. $\tan\left(\theta\right)=x/z$, $\nu=0.214$, $\lambda=635\rm nm$. The black lines in the bottom half depict several ray trajectories, Eq.\ (\ref{eq:path}). Upper half shows trajectories for $8\times$ enhanced lens-strength, i.e. $\nu=1.71$.\label{Schršdi2DImage}}
\end{figure}

On the positive $z$-axis the wave-function Eq.\ (\ref{SchroediExactSolution}) satisfies\cite{LandauQM,Deguchi1986} $|\Psi_C\left(z\right)|^2=2\pi\nu/\left[e^{2\pi\nu}-1\right]$. The amplitude of this wave function is shown in Fig.\ (\ref{Schršdi2DImage}). 

Now we will write down the scalar Helmholtz-Equation for light \cite{SalehTeich}, $\nabla^2 U+k^2 [n\left(\mathbf{r}\right)^2\!\! /n_0^2 ]\,U=0$, with the refractive index profile given by Eq.\ (\ref{eqn:RefracGradient}). One should think of this equation as being the analogon to the SE with non-zero energy $E=n_0^2/2$ and potential energy $V=-n\left(\mathbf{r}\right)^2\!\!/2 + n_0^2/2$ as before in Section \ref{SectionPhotRutherford}. We find:
\begin{equation}
\nabla^2 U+k^2 \left[1 + \frac{2\Delta n R}{n_0 r}+\left\{\frac{\Delta n^2 R^2}{n_0^2 r^2}\right\}\right]U=0.
\end{equation}
A comparison with Eq.\ (\ref{SchroediEq}) allows the identification of the interaction parameter $\nu$ as
\begin{equation}
\nu \rightarrow -k\frac{\Delta n R}{n_0}=\frac{k}{\xi},\label{nuIdentification}
\end{equation}
to first order in the small quantity $\Delta n/n_0\ll 1$. Infact, this identification could have been guessed without this inspection simply by the definition of the parameter $\nu=\frac{C k}{2E}$ and the classical correspondences found earlier with its prescription $2E/C \rightarrow \xi$. The solution $\Psi_C$, Eq.\ (\ref{SchroediExactSolution}), to the SE of the Coulomb scattering problem may thus be used to find the scalar field amplitude $U$ in the case of diffraction by the inhomogeneous refractive index field Eq.\ (\ref{eqn:RefracGradient}). Again, the particle problem may thus be used to obtain results for its corresponding optical phenomenon, similar to the mechanical-optical analogon which is the "F=ma"-optics framework. The strength- and polarity-encoding parameter $\nu$ is found to be proportional to $\propto \xi^{-1}$, which was the parameter describing the ray-trajectories.

Another view on this equivalence is obtained by looking at the Kirchhoff diffraction for this refractive index profile. We have recently used the Fresnel-grade approximation of the diffraction on this refractive index profile to describe the photothermal signal of single heated nano-particles \cite{SelmkeDiffraction}. We may write for the scalar field amplitude $U$ in the image-plane located at a distance $z$ behind the aperture-plane the following diffraction integral:
\begin{equation}
U=\frac{k}{iz}\, e^{ikz+i\frac{kx^2}{2z}}\!\! \int_{0}^{\infty} \!\!\! \! U_a\, e^{i \frac{k\rho^2}{2z}} J_0\left(\frac{k\rho x}{z}\right)\exp\left(i \Delta \chi_\rho\right)\rho\,\mathrm{d}\rho.
\end{equation}
Here, $U_a$ is the field in the aperture-plane at $z=0$. The collected phase advance $\Delta \chi_\rho$ may be computed in a straight-ray approximation to yield, neglecting an additional constant phase,
\begin{equation}
\Delta \chi_\rho \left(\rho\right) =k_0\!\int_{-l}^{l} \!n\left(\!\sqrt{z^2+ \rho^2}\right) \mathrm{d} z \approx -2k_0 R\,\Delta n \ln\left(\frac{\rho}{2l}\right)\nonumber \label{EqPhaseAdvance},
\end{equation}
and is caused by a travelled distance $l$ both in front and behind the lens. To mimic the quantum mechanical initial state of a plane wave used before, we will consider a unit amplitude plane wave in the aperture plane, i.e.\ $U_a=1$. The abbreviations $k=k_0 n_0$, $a=\Delta n R/n_0$ and $\zeta=-\frac{ik}{2z}$ are used from now on. We will also use the relations ${}_{1}F_{1}\left(a;b;-x\right)=e^{-x}{}_{1}F_{1}\left(b-a;b;x\right)$, $\log\left(-i\right)=-i\pi/2$. While in our previous paper we have studied the limit of $z \ll k \omega^2$, we will now retain the $z$-dependence. Using the symbolic arithmetics program \textit{Mathematica}, or the integral representation of the confluent hypergeomeric function ${}_{1}F_{1}$ in combination with the Bessel function representation through ${}_{0}F_{1}$, one finds the image-plane amplitude to equal
\begin{widetext}
\begin{eqnarray}
&U\left(x,z\right)=\frac{k}{zi}e^{ikz+i\frac{kx^2}{2z}}\frac{1}{2}\zeta^{-1+ika}\Gamma\left(1-ika\right){}_{1}F_{1}\left(1-ika;1;-\frac{k^2 x^2}{4z^2\zeta}\right)\label{DiffractionFocused}\\
&= e^{ika\log\left(\frac{k}{2z}\right)}e^{ka\frac{\pi}{2}}e^{ikz}\,\Gamma\left(1-ika\right){}_{1}F_{1}\left(ika;1;i\frac{kx^2}{2z}\right)\label{EqnFresnelPW}
\end{eqnarray}
\end{widetext}
Which already resembles the exact solution $\Psi_C$ found earlier. If one considers the argument of the hypergeometric function in the forward direction using the relation $x^2=r^2-z^2\approx 2z\left[r-z\right]$, one may write $i\frac{kx^2}{2z}\approx ik\left[r-z\right]$, such that an agreement in form is found between Eq.\ (\ref{EqnFresnelPW}) and Eq.\ (\ref{SchroediExactSolution}) up to a logarithmic phase-factor. Upon inspection of these two equations, we may thus also arrive at the identification $\nu \rightarrow -ka$, which is the same as stated in Eq.\ (\ref{nuIdentification}).
The absolute value of the final expression for the field amplitude shows that for $\Delta n=a=0$ one obtains $U\!\left(\mathbf{r}\right)=e^{ikz}$, i.e.\ the plane wave emerges unperturbed (${}_{1}F_{1}\left(a;b;x\right)=1$ as $x\rightarrow 0$\cite{AbramowitzStegun}).
\begin{figure}[ht]
\begin{center}\includegraphics [width=1.0\columnwidth]{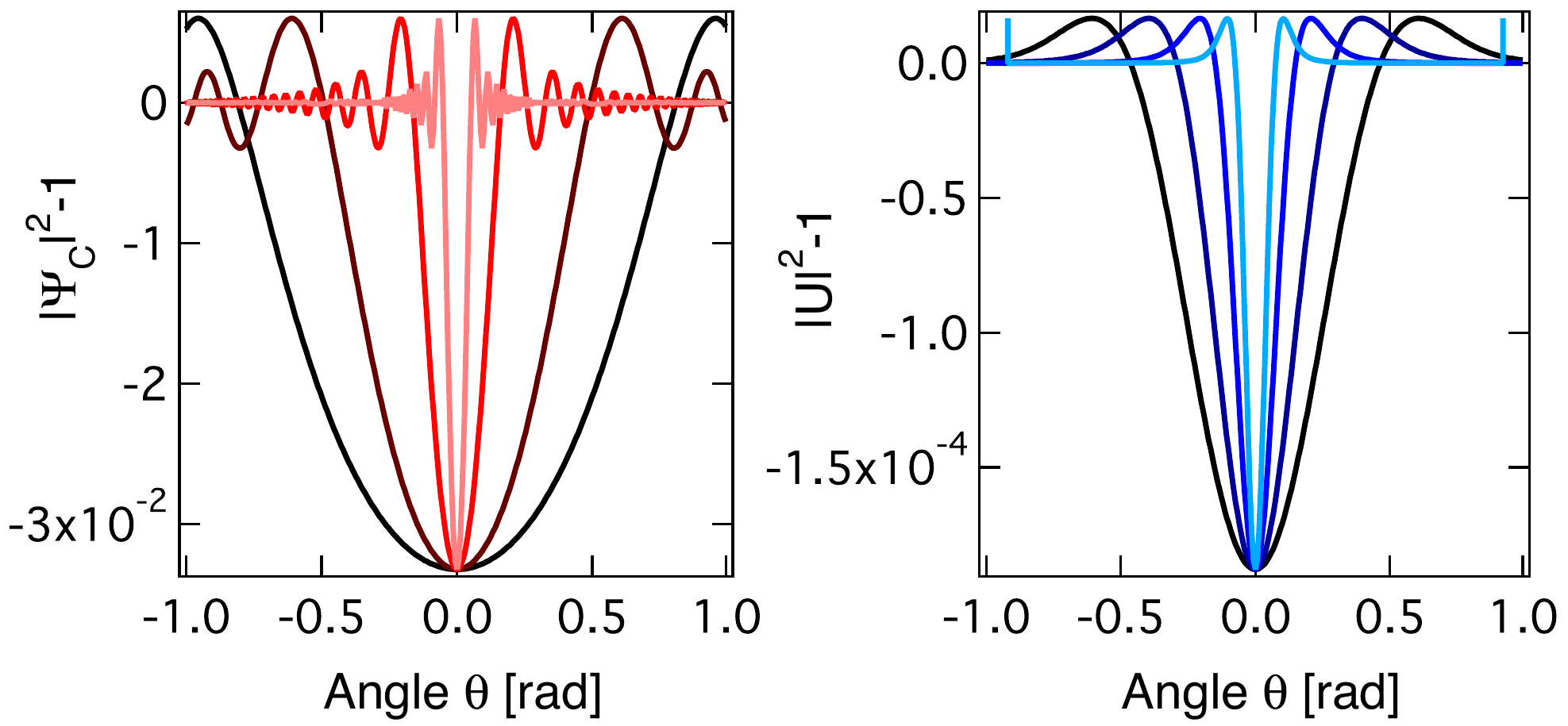}
  \end{center}
\caption{Left: $|\Psi_C\left(\theta\right)|^2$ from Eq.\ (\ref{SchroediExactSolution}) at $z=\left\{300,10^3,10^4,10^5\right\}\rm nm$ (black to light red). $\nu=0.0106$. Right: $\left[|U|^2-1\right]/|U|_{\theta=0}^2$ Eq.\ (\ref{DiffractionFocused}) for $\omega=\left\{300,500,1000,2000\right\}\rm nm$ (black to cyan).\label{FigPsiScans}}
\end{figure}
We have thus demonstrated the equivalence of the plane-wave (pw) quantum mechanical Coulomb scattering problem and the pw diffraction by our specific thermal lens $n\left(r\right)$, to first order $\mathcal{O}\left(\Delta n/n\right)$. 

The connection shall now be demonstrated between these wave-mechanical descriptions and the previously studied classical cases. While the shape of the wave-function amplitude already resembles the family of trajectories of a given energy but varying impact parameter, i.e.\ Fig.\ \ref{fig:scat}b, the resemblance is misleading. For a given constant wavelength $\lambda$ and thus constant wavenumber, the diffraction pattern is independent on the magnitude of the thermal lens, i.e.\ $\Delta n$, while the family of trajectories would change. For plane wave scattering/illumination the spatial features and patterns of the perturbed wave-amplitude do not resemble the trajectories and shadows in the near field as predicted by geometrical optics\cite{Zakowicz2002}. While classical dynamics and scattering descriptions require the notion of paths and trajectories, in the wave-mechanical scattering description no clear correspondence exists for the case of plane wave or wide beam scattering\cite{Zakowicz2002Ext}. It is only in the far field that the classical average particle number density coincides, up to an additional zero-mean oscillation with an undetectably high spatial frequency , with the QM-wavefunction amplitude\cite{Samengo1999} and thus with the classical expressions for the cross-section $\left(\mathrm{d}\sigma/\mathrm{d}\Omega\right)$. We will therefore now formulate the correct limit which connects both wave and the classical descriptions.

\subsection{QM: Wave Packet Scattering\label{SectionWPScattering}}
In order to draw the connection between the classical and the wave pictures both in optics and quantum mechanics as we have described above, it is necessary to reconsider what constitutes this classical limit such that a recovery may be demanded. In the previous paragraph it was already shown that the plane-wave approach does not resemble the classical pictures apart from the total far-field scattering-cross-section. In optics the transition is reached by letting the wavelength go to zero, $\lambda\rightarrow 0$. Then, the wave-front normals will follow the trajectories described by ray-optics \cite{BornWolf1980Book}. Due to the possession of the exact solution in the case of quantum mechanical wave theory, its transition to the classical particle trajectory picture will be outlined here. Here, we will investigate the QM scattering problem of a wave-packet (wp) and will afterwards instead of following phase-front normals quantify the scattering of a specific wave-packet which will resemble a confined minimally spreading ray. Following Baryshevskii et al. \cite{Baryshevskii2004} and similar works\cite{Skoromnik2010,Zakowicz2003,Zakowicz2009}, one may write an initial wave packet localized near (i.e.\ focused at) $\mathbf{r_0}$ at time $t=0$ as:
\begin{eqnarray}
\Psi_0^{\rm wp}\left(\mathbf{r},0\right)&=&\int\mathrm{d}\mathbf{k}\, A\left(\mathbf{k}\right) e^{i \mathbf{k}\cdot\left(\mathbf{r}-\mathbf{r_0}\right)}\\
A\left(\mathbf{k}\right)&=&\int \mathrm{d}\mathbf{r}\, G\left(\mathbf{r}\right)e^{-i \mathbf{r}\cdot\left(\mathbf{k}-\mathbf{k_0}\right)}.
\end{eqnarray}
Such a superposition may even have the familiar phase anomaly known from wave optics as the Gouy phase \cite{Wolf2010Gouy}. The functions $A\left(\mathbf{k}\right)$ and $G\left(\mathbf{r}\right)$ define the wave-packet form in momentum- and real-space, respectively.
For different momenta $\mathbf{k}$ and thus possibly different interaction parameters $\nu_k=k\frac{C}{2E}$ or $\nu_k=k\xi^{-1}$ the solution formerly written down for $\mathbf{k}=k\, \mathbf{\hat{z}}$, Eq.\ (\ref{SchroediExactSolution}), will now be given in a fixed coordinate frame for arbitrary direction of the incident wave-vector:
\begin{figure}[h!t]
\begin{center}\includegraphics [width=0.8\columnwidth]{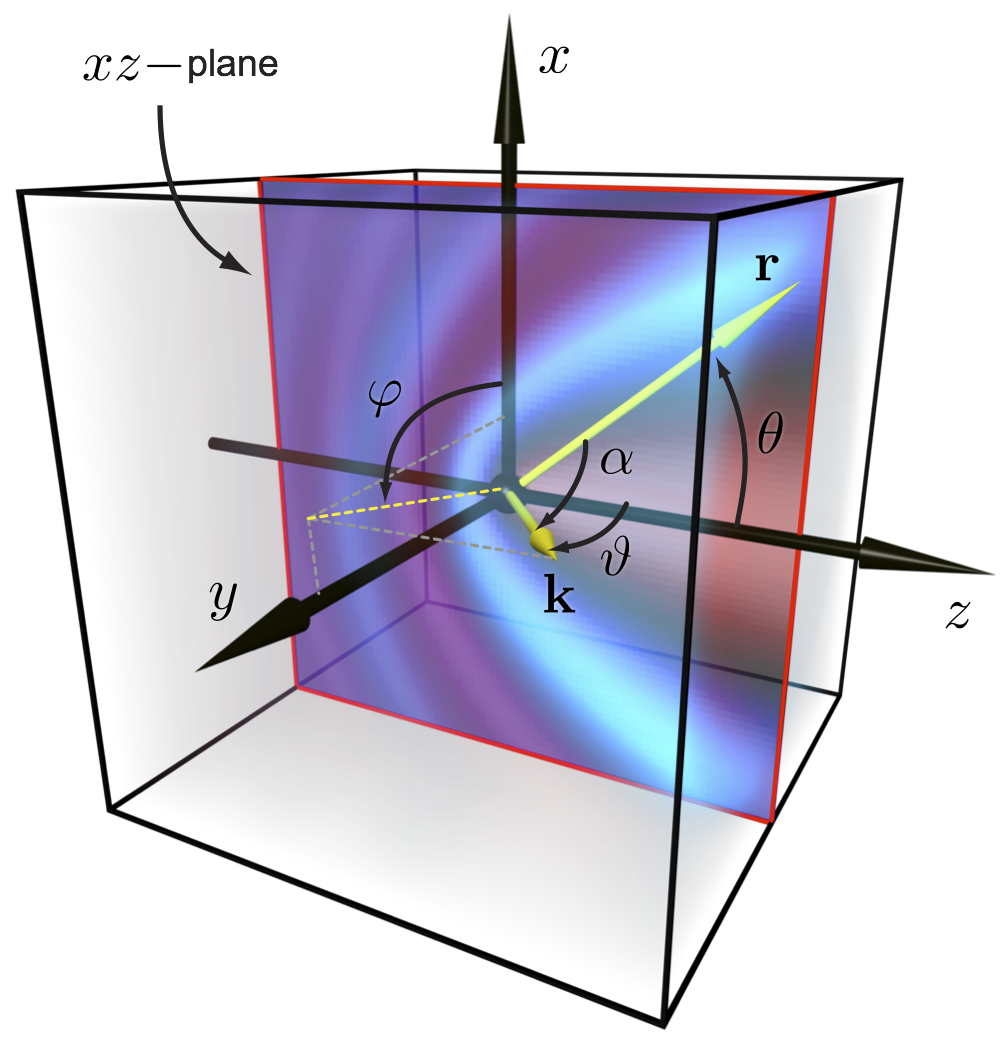}
  \end{center}
\caption{Geometry for wave-vector $\mathbf{k}$ of $\Psi^{\mathbf{k}}_C\left(\mathbf{r}\right)$ used in the wave-packet superposition $\Psi^{\rm wp}_C\left(\mathrm{r},t\right)$. The azimuthal angle of the $xz$-plane is $\phi=0$ for $x\ge 0$ and $\phi=\pi$ for $x<0$, while the polar angle is $\theta=\arccos\left(z/r\right)$ with $r^2=x^2+z^2$.\label{FigQMGeom}}
\end{figure}
\begin{figure}[h!t]
\begin{center}\includegraphics [width=1.0\columnwidth]{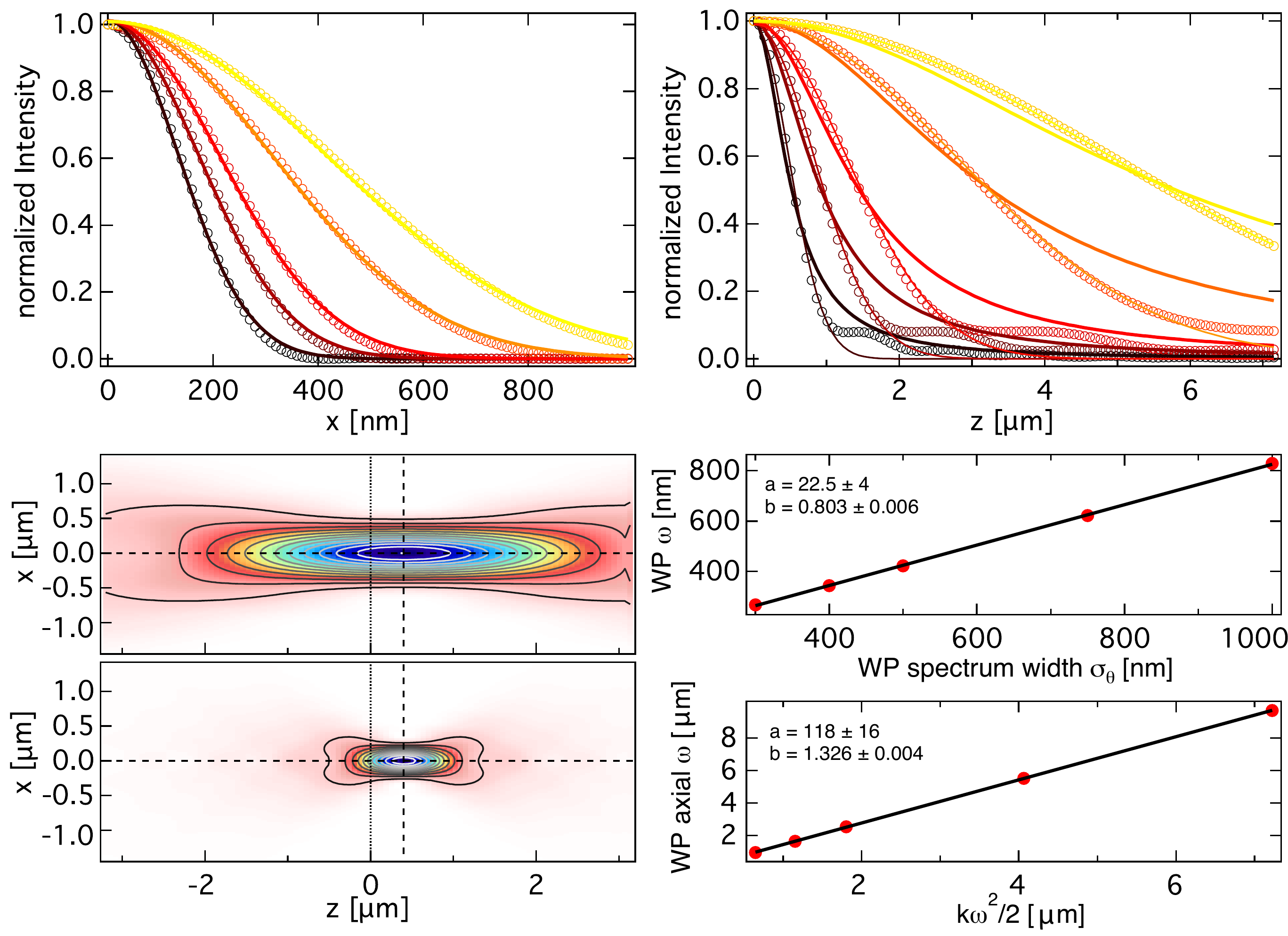}
  \end{center}
\caption{Initial wave-packet amplitudes $|\Psi_0^{\rm wp}\left(\mathbf{r}\right)|^2$. Images show two focused WPs with $\mathbf{z_0}=+400{\rm nm}\,\mathbf{\hat{z}}$ and $\sigma_\vartheta=\left\{15^{\circ},30^{\circ}\right\}$ degrees. Axial Gaussian fits yield $\omega_0=23{\rm nm} + 0.8 \omega_\vartheta$ with $\omega_\vartheta=2/\!\left[k\sigma_\vartheta\right]$ and axial Gaussian widths $\omega_{0,z}=120{\rm nm} + 1.32 z_R$ with $z_R=k \omega_\vartheta^2 /2$.\label{FigWP02}}
\end{figure}
\begin{equation}
\Psi^{\mathbf{k}}_C\left(\mathbf{r}\right)=e^{-\frac{\pi}{2}\nu_k}e^{i\mathbf{k}\cdot\mathbf{r}}\Gamma\left(1+i\nu_k\right)\!{}_{1}F_{1}\left(-i\nu_k;1;i\left(kr\!-\!\mathbf{k}\!\cdot\!\mathbf{r}\right)\right).\label{SchroediExactSolutionk}
\end{equation}
The time evolution of an arbitrary initial wave-packet as described by Eq.\ (\ref{eq:WP0}) will be determined by the superposition $\Psi^{\rm wp}_C$ of the individual plane-wave solutions corresponding to the pw-spectrum components of this initial wave-packet:
\begin{equation}
\Psi^{\rm wp}_C\left(\mathrm{r},t\right)=\int\!\mathrm{d}\mathbf{k}\,A\left(\mathbf{k}\right) e^{-i \mathbf{k}\cdot\mathbf{r_0}}\, \Psi^{\mathbf{k}}_C\left(\mathbf{r}\right) e^{-i\frac{\hbar k^2}{2m}t}\label{eq:TDSEgeneral}
\end{equation}
Now, assuming that only different incident angles will contribute, we have a constant momentum magnitude of $|\mathbf{k}|=\bar{k}$. Further, we will consider an azimuthally symmetric angle distribution for the wave-vector spectrum such that it takes the form $A\left(\mathbf{k}\right)=A\left(k,\vartheta,\varphi\right)=\delta\left(\bar{k}-k\right)A\left(\vartheta\right)/2\pi$, where the wave-vector has been express in spherical coordinates. The initial wave-packet then reads:
\begin{eqnarray}
\Psi_0^{\rm wp}\left(\mathbf{r},0\right)&=&\int_{0}^{\pi}\mathrm{d}\vartheta\, A\left(\vartheta\right) e^{-i \bar{k} \left[z_0-r\cos\left(\theta\right)\right]\cos\left(\vartheta\right)} \label{eq:WP0}\\
&&\quad  \times \sin\left(\vartheta\right) J_0\left(\bar{k} r \sin\left(\theta\right)\sin\left(\vartheta\right)\right)\nonumber
\end{eqnarray}
More specifically, we will choose the polar-angle spectrum $A\left(\vartheta\right)$ to be a Gaussian with an angular width of $\sigma_\vartheta$,

\begin{equation}
A\left(\vartheta\right)=\exp\left(-\frac{\vartheta^2}{2\sigma_\vartheta^2}\right),\label{eq:WP0Specific}
\end{equation}
which results in a focused wave-packet that has similar properties as a TEM00-mode Gaussian beam with a characteristic width-scale given by $\omega_\vartheta=2/\!\left[k\sigma_\vartheta\right]$ (see Fig.\ \ref{FigWP02}). This immediately implies that the angular spreading of the WP decreases, i.e.\ becomes paraxial and resembling a ray, when the wavelength decreases since $\sigma_\vartheta\propto \lambda/\omega_\vartheta$. In fact, as is the case for the beam emerging from a laser-pointer, its lateral intensity follows closely a Gaussian distribution of width $\omega_0$, i.e.\ $|\Psi_0^{\rm wp}\left(x\right)|^2\propto \exp\left(-2x^2/\omega_0^2\right)$ while the axial intensity pattern is enveloped by a Lorenzian profile with a corresponding of range $z_R=k \omega_0^2/2$, i.e.\ $|\Psi_0^{\rm wp}\left(z\right)|^2\propto 1/\left[1+z^2/z_R^2\right]$ (solid thick lines). A fit in axial direction may also yield some axial Gaussian with widths $\omega_{0,z}$ which fit the central bumps (thin solid lines). The phase-pattern also shows the Gouy-phase anomaly, i.e.\ a phase advance of $\pi$ as compared to a spherical wave emanating from the focus.

The solution to the time-dependent SE, Eq.\ (\ref{eq:TDSEgeneral}), with the specific choice of the initial wave-packet as described by Eq.\ (\ref{eq:WP0Specific}), then reads:
\begin{widetext}
\begin{equation}
\Psi^{\rm wp}_C\left(\mathrm{r},t\right)=\!\int_{0}^{\Theta}\!\!\!\mathrm{d}\vartheta\int_{0}^{2\pi}\!\!\!\!\mathrm{d} \varphi \,\sin\!\left(\vartheta\right)\!A\left(\vartheta\right) e^{-i \bar{k} z_0 \cos\left(\vartheta\right)} e^{-\frac{\pi}{2}\nu}e^{i\mathbf{k}\cdot\mathbf{r}}\Gamma\left(1+i\nu\right){}_{1}F_{1}\left(-i\nu;1;i\left(\bar{k} r-\mathbf{k}\cdot\mathbf{r}\right)\right) e^{-i\frac{\hbar \bar{k}^2}{2m}t}
\end{equation}
\end{widetext}
with the dot product of the wave-vector and the radius vector in spherical coordinates $\mathbf{k}\cdot\mathbf{r}=\bar{k}\, r\cos\left(\alpha\right)=\bar{k}\, r \left[\cos\left(\theta\right) \cos\left(\vartheta\right)+ \sin\left(\theta\right) \sin\left(\vartheta\right)\cos\left(\phi-\varphi\right)\right]$ (see Fig.\ \ref{FigQMGeom}), $\nu_k=\nu$ and $\mathbf{k}\cdot\mathbf{r_0}=\bar{k}\, z_0 \cos\left(\vartheta\right)$. 

\begin{figure}[h!t]
\begin{center}\includegraphics [width=1.0\columnwidth]{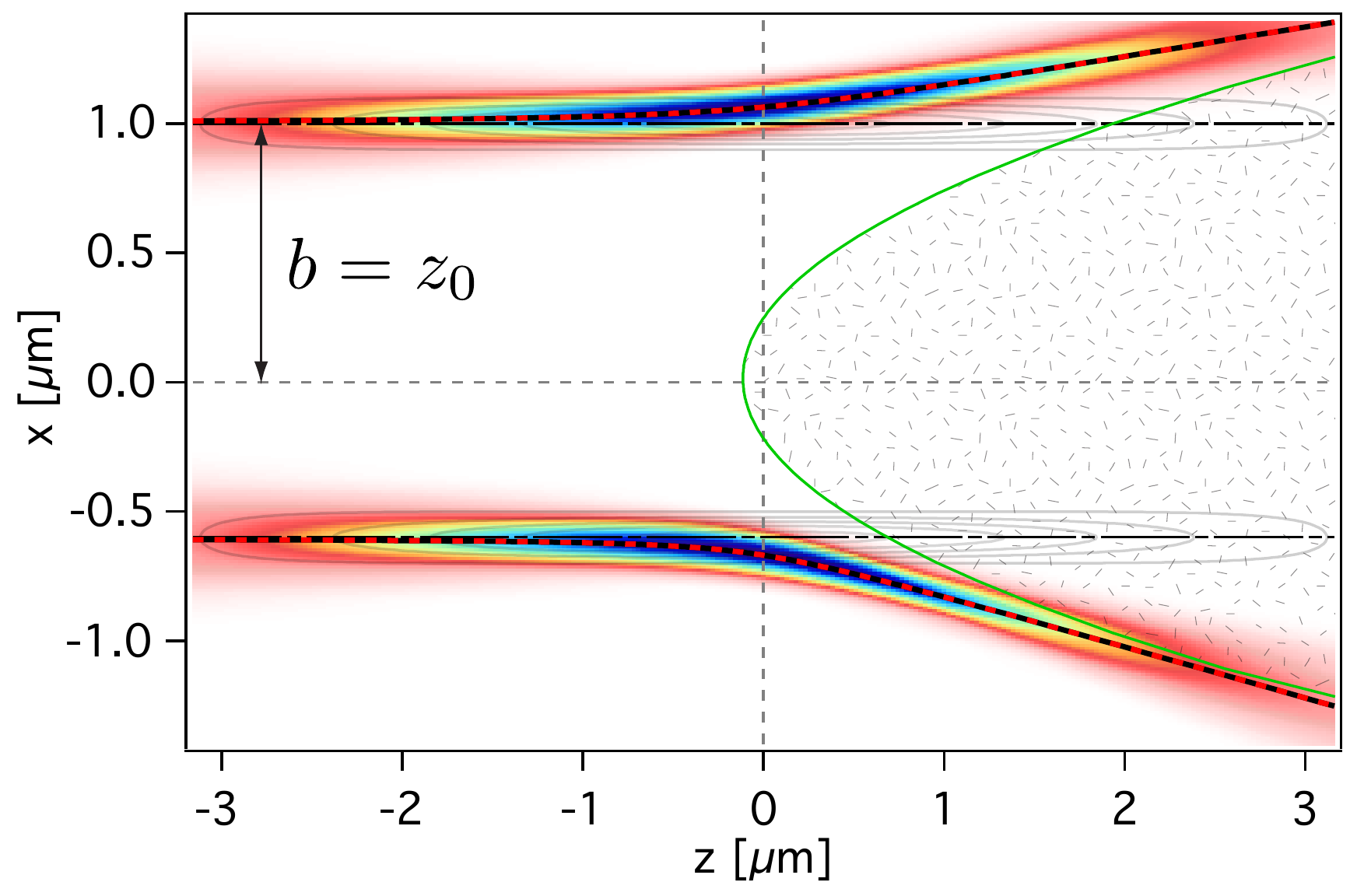}
  \end{center}
\caption{Initial wave-packet amplitudes $|\Psi_0^{\rm wp}\left(\mathbf{r}\right)|^2$. Image shows two focused WPs with $\mathbf{z_0}=-0.600\mu{\rm m}\,\mathbf{\hat{x}}$ and $\mathbf{z_0}=1.0\mu{\rm m}\,\mathbf{\hat{x}}$ with a spreading of $\sigma_\vartheta=3^{\circ}$ degrees and a corresponding width-scale of $\omega_\vartheta=0.13\mu{\rm m}$. The strength of the potential was $\nu=17.1$, the wave-number was $\bar{k}=289\mu{\rm m}^{-1}$. The scattered WPs follow the classical photonic Rutherford trajectories (red-black dashed), similar to Fig.\ \ref{fig:scat}b, and avoid the shadow region (green line, textured area).\label{FigWPRF}}
\end{figure}

To obtain the ray-limit, we finally choose beams of a finite width $\omega_\vartheta$ and small angular spread (i.e.\ paraxial) with an lateral offset in $x$-direction of the resulting stretched wave-packet to some $z_0 \gg \omega_\vartheta$. In this case one must set $\mathbf{k}\cdot\mathbf{r_0}=\bar{k}\, z_0 \sin\left(\vartheta\right)\cos\left(\pi-\varphi\right)$ (see Fig.\ \ref{FigQMGeom}). We then find the wave-packets to be distorted by the scattering process such that its probability amplitude $|\Psi^{\rm wp}_C\left(\mathrm{r}\right)|^2$ follows the classical Rutherford scattering trajectory $r\left(\theta\right)$, Eq.\ (\ref{eq:Rutherford_Photon}) with the plane polar angle $\phi$ now being the polar angle $\theta$, with the impact parameter set to the initial WP lateral offset $b\rightarrow z_0$ and the lens strength parameter $\xi\rightarrow \bar{k}/\nu$. This correspondence is shown in Fig. \ref{FigWPRF} and is the analogue to Fig.\ \ref{fig:scat}b. This is the expected classical property of the quantum mechanical scattering description\cite{Zakowicz2003}. 

\begin{figure}[h!t]
\begin{center}\includegraphics [width=1.0\columnwidth]{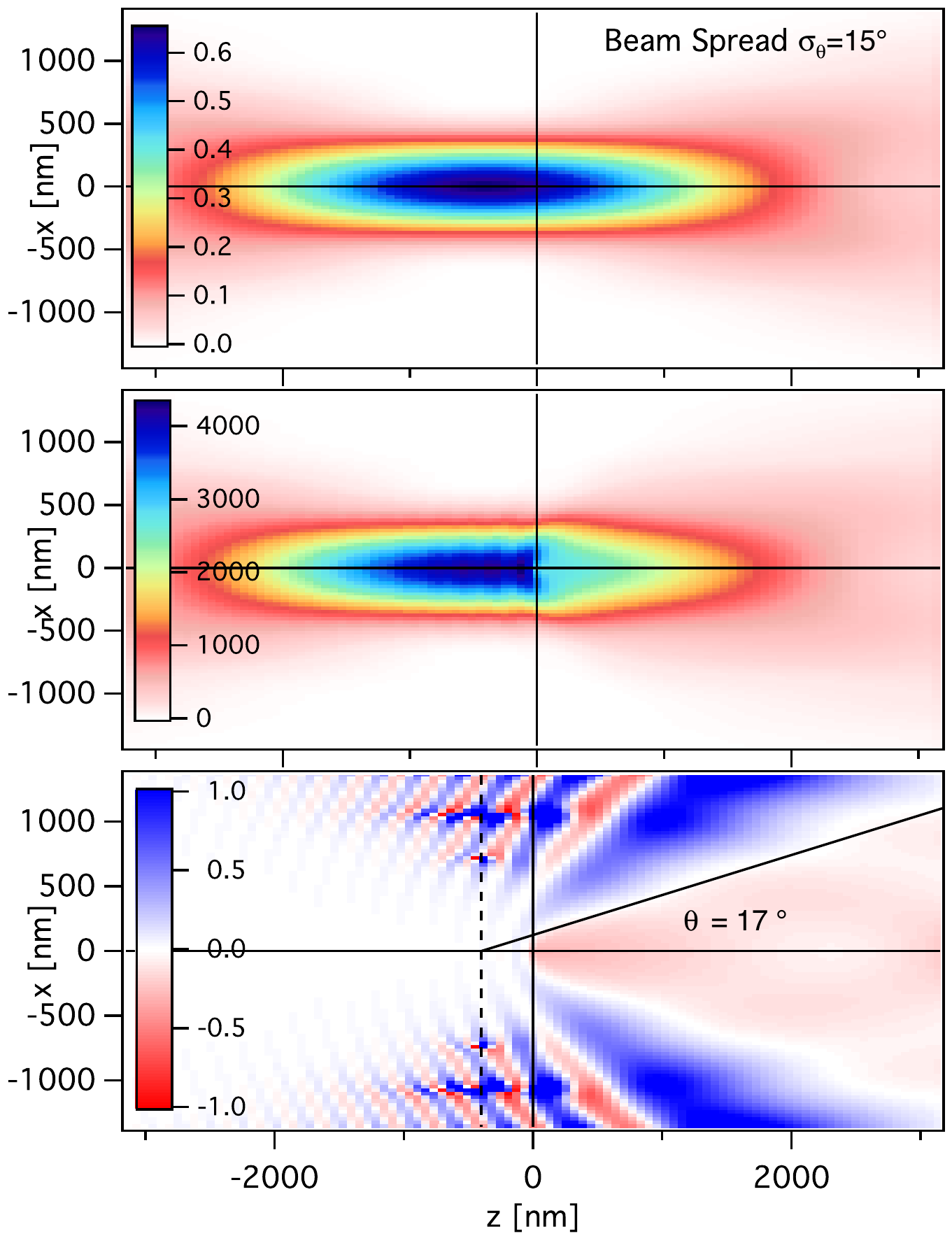}
  \end{center}
\caption{Focused wave-packet scattering on a Coulomb potential with no axial offset of the initial wave-packet and finite spreading width $\sigma_\vartheta=15^{\circ}$. Depicted are $|\Psi^{\rm wp}_0|^2$ (top), $|\Psi^{\rm wp}_C|^2$ (center) and their normalized difference (bottom). \label{FigPTDiffraction}}
\end{figure}

Having previously established the connection to the optical wave-mechanical framework, one may also think of these trajectories to be refracted narrow gaussian beams. The power of the general beam description is shown in the following via its application to photothermal microscopy (see also Fig.\ \ref{FigPTDiffraction}).
\subsection{Perturbation quantification: The normalized difference / photothermal signal $\Phi$}
In a photothermal microscopy experiment a focused beam is used to probe the refractive index lens of a heated nano particle. A Gaussian beam provided by a TEM00-mode operated laser will be focused by a microscope objective. It will thus have a total angular spread\cite{SalehTeich} of twice the divergence half-angle $\theta_{\rm div}=\frac{2}{k\omega}$. For $\lambda=635$ in a medium with $n_0=1.46$ and a beam waist $\omega_0=281\,{\rm nm}$ one finds $\theta_{\rm div}\approx 28^{\circ}$. Therefore, one needs to consider both wave-mechanical pictures with focusing.
First, we will discuss the QM analog on to the rel.\ PT signal. To this end we will need to focus on the narrow forward direction interference-zone which is otherwise neglected in standard treatments of Coulomb scattering problem as its angular extent shrinks to zero at large distances. However, as it will turn out, it is exactly this interference which causes the PT signal and the interference-zone is expanded to detectable angular extents via the finite-width plane-wave wave-vector spectrum. In the small angle interference domain (see Fig.\ (\ref{FigPsiScans})), we need the asymptotic form of $\Psi_C\left(r,z\right)$, Eq.\ (\ref{SchroediExactSolution}), for small $\eta=r-z$. As stated before, in the treatment of QM Coulomb scattering the opposite limit of $k\eta\rightarrow \infty$ is usually considered\cite{Grama2002}. Now, series expansion of the confluent hypergeometric function reads $_1 F_1\left(a;b;z\right)=1+\frac{a}{b}z+\mathcal{O}\left(z^2\right)$. Using $e^x\approx 1+x$ for small $x\ll 1$, Eq.\ (\ref{SchroediExactSolution}) thus becomes to order $\mathcal{O}\left(\eta\right)$ and to order $\mathcal{O}\left(\nu\right)$: 
\begin{eqnarray}
\Psi_C\left(\mathbf{r}\right)&\approx& \left[1 -\frac{\pi}{2}\nu\right] e^{ikz}\left[1-i\gamma_E \nu\right]\left[1-i\nu \times ik\eta\right]\\
&\approx& e^{ikz}\left\{1+ \nu \left[-\frac{\pi}{2} + k\eta\right] + i \nu \left[-\gamma_E + \frac{k^2\eta^2}{4}\right]\right\}\nonumber
\end{eqnarray}
since $\eta=r\left[1-\cos\left(\theta\right)\right]=2r\sin^2\left(\theta/2\right)$ we have to first order in $\nu$ again the squared modulus of the wave-function:
\begin{equation}
|\Psi_C\left(\theta,r\right)|^2\approx1+\nu\left[4 k r \sin^2\left(\theta/2\right) - \pi\right]
\end{equation}
This approximation nicely fits the central bump. For zero angle, i.e.\ in forward direction, the finite value $|\Psi_C\left(\theta=0\right)|^2=1-\pi \nu$. A more rigorous demonstration of this limit can be found in loc.\ Eq.\ (42) of reference \cite{Deguchi1986}.
The Photothermal signal analogue would thus read, for plane-wave illumination and small angles:
\begin{equation}
\Phi=|\Psi_C|^2-1 \approx -\pi\nu + k r \nu\, \theta^2 \label{PTQMinterference}
\end{equation}
It is a parabola which cuts the $|\Psi_C|^2=0$-axis at $\theta_0=\pm \sqrt{\frac{\pi}{kr}}$ independent of the strength of the perturbation $\xi$. This is the width of the central bump and defines the "near"-zone as in loc.\ Eq.\ (9) of reference \cite{Baryshevskii2004} derived on ground of different arguments. It depends on the distance $r$ to the scatterer. Since analytical progress in the case of wave-packet scattering was here non-feasable, we will now concentrate on the corresponding optical scenario to achieve a more general expression which will generalize Eq.\ (\ref{PTQMinterference}) to the non plane-wave case.

If one assumes a Gaussian beam which illuminates the aperture plane a corresponding substitution of $U_a$ by
\begin{equation}
U_a=\frac{\omega_0}{\omega\left(z_0\right)}\exp\left(-\frac{\rho^2}{\omega^2\left(z_0\right)}+ikz_0+i\frac{k\rho^2}{2R_C\left(z_0\right)}-i\zeta_G\right)\nonumber\label{eq:GaussU}
\end{equation}
into Eq.\ (\ref{DiffractionFocused}) requires the substitution of $\zeta=\frac{1}{\omega\left(z_0\right)^2}-\frac{ik}{2z}-\frac{ik}{2R_C\left(z_0\right)}$ and a pre-factor including a phase factor (Gouy-phase) and an amplitude factor $\omega_0^2/\omega\left(z_0\right)^2$. The beam-waist of the focused Gaussian beam is denoted by $\omega\left(z_0\right)^2=\omega_0^2\left[1+z_0^2/z_R^2\right]^2$ and the local radius of curvature by $R_C\left(z_0\right)=z_0\left[1+z_R^2/z_0^2\right]$ and the Gouy-phase is $\zeta_G=\arctan\left(z_0/z_R\right)$.

In the case of the diffraction formulation we can write for the relative photothermal signal $\Phi$ on the $z$-axis in case of an illumination Gaussian beam from Eq.\ (\ref{DiffractionFocused}):
\begin{equation}
\Phi = \frac{|U_\nu|^2 - |U_{\nu=0}|^2}{|U_{\nu=0}|^2} = e^{2\nu\arg\left(\zeta\right)}|\Gamma\left(1+i\nu\right)|^2-1
\end{equation}
The number $|\nu|=k R|\Delta n|\ll 1$ is small, such that $\Gamma\left(1+i\nu\right)\approx 1-i\gamma_E \nu+\mathcal{O}\left(\nu^2\right)$ where $\gamma_E$ is Euler's constant. This means that to first order in $\mathcal{O}\left(\nu\right)$ one may write 
\begin{equation}
\Phi = 2\nu\arg\left(\zeta\right)= 2\nu\arctan\left(\frac{-z_0}{z_R}\right)\label{eqPhiApproxDiffrOnAxis}
\end{equation}
The plane-wave limit may be read off directly from Eq.\ (\ref{EqnFresnelPW}), $\Phi = -\pi\nu$. It also agrees in value with the above Eq.\ (\ref{eqPhiApproxDiffrOnAxis}) for large offsets of the beam-waist, i.e.\ $|z_0|\gg z_R$ and $z_0>0$ (scatterer behind beam-waist). It is clear by the discussions and equations presented so far, that photo thermal single particle microscopy as described in references \cite{SelmkeDiffraction,Selmke2012ACSNano} deals with the interference zone encountered in Section \ref{SubSectionPWDiffraction} of plane-wave quantum mechanical Coulomb / Rutherford scattering. In this case the interference zone exhibited a vanishing angular extent $\theta_0=\pm \sqrt{\frac{\pi}{kr}}$ thus being typically discard in scattering analysis and leading to the $\sin^{-4}$-dependence of the detectable cross-section. In photothermal microscopy it is modified and widened up to the extend of the angular spread of plane-wave contributions which make up the incident beam, i.e.\ the angular spread of the probing laser-beam. Thereby, the interference zone corresponds to the detected angular domain of photothermal microscopy and determines the signal. On the other hand, if a deflection is measured of the probing beam, the usual low-energy (wide lateral wave-packet width) limit of Rutherford scattering is attained and deflections by the scattering angle $\theta$ are expected in the limit of small angular spread and far enough lateral offsets (see discussion in Section \ref{SectionWPScattering}).

\begin{figure}[h!t]
\begin{center}\includegraphics [width=1.0\columnwidth]{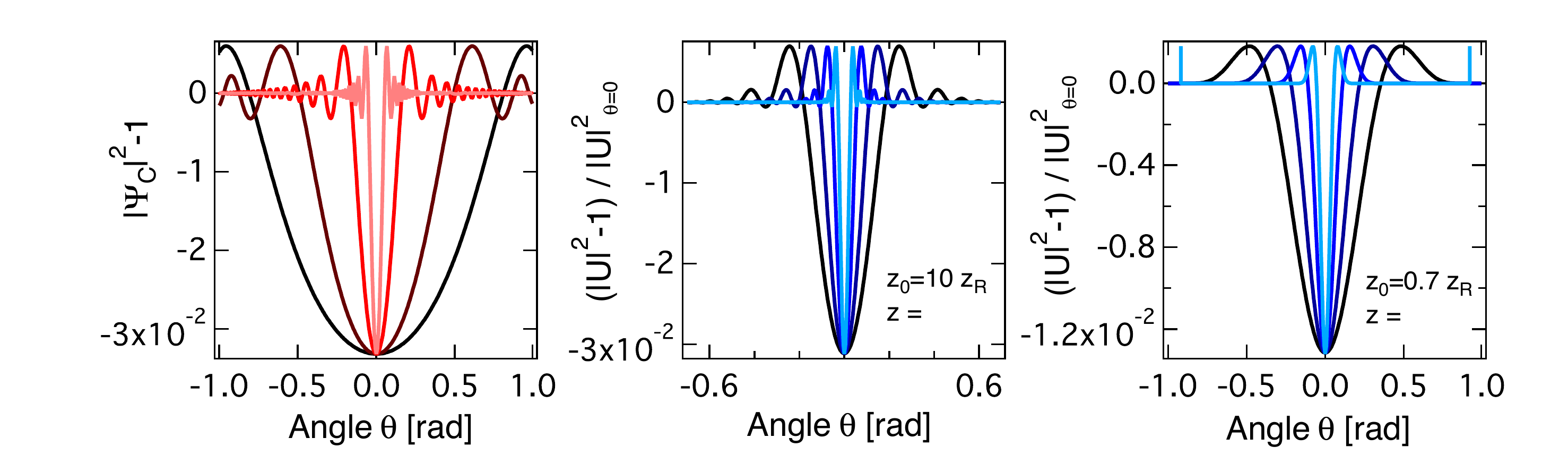}
  \end{center}
\caption{QM wave mechanical probability amplitude $\Psi_C\left(x\right)$, Eq.\ (\ref{SchroediExactSolution}) at various axial coordinates $z$ (red) and wave-optical scalar diffraction results, Eq.\ (\ref{DiffractionFocused}), for various beam waists (blue). The optical diffraction results have been computed using a Gaussian beam illuminating the aperture-plane as described by Eq.\ (\ref{eq:GaussU}).}
\end{figure}

\newpage
\section{Appendix}
\subsection{Binet's equation for pertubative forces}
While the perturbative force here is $\propto r^{-3}$, the perturbative force as obtained from the geodesic equation\cite{Eliseo2006} in the Schwarzschild metric of a massive body which enters the gravitational two-body problem is $\propto r^{-4}$. Given below are the differential orbit equations for the perturbed motion of massive particles in the potential fields $V\left(r\right)=C r^{-1}+\alpha r^{-2}+\beta r^{-3}$. From Binet's equation $F\left(u\right)=-m L_z^2 u^2\left[\partial^2_{\phi}u+u\right]$ one readily finds
\begin{equation}
\partial^2_{\phi}u + u\left[1 - \frac{2\alpha}{m L_z^2}\right] - u^2\frac{3\beta}{m L_z^2}=\frac{C}{m L_z^2}
\end{equation}
While $\left\{\alpha\ne 0,\beta=0\right\}$ corresponds to Fermat's least optical path Eq.\ (\ref{eq:u}), the combination $\left\{\alpha= 0,\beta\ne 0\right\}$ matches in form the equation of motion of a massive particle in the Schwarzschild metric. The null geodesic equation for light rays requires to put $\frac{C}{m L_z^2}=0$ in the latter one and will thus also not correspond to the path described by Eq.\ (\ref{eq:u}).\cite{Evans1996GRTphotpart,Hemzal2006GRTThesis}
\subsection{Alternate trajectory formulation}
An equivalent trajectory formulation similar to the one given in reference \cite{Samengo1999}. The distance $r_C$ of closest approach for a repulsive potential for a head-on impact, i.e. $b=0$, will be used. In the mechanical case, this may be evaluated by setting the kinetic energy of the incoming particle at infinity equal to the potential energy at the minimum distance, $E=\frac{C}{r_C}$, yielding $r_C=C/E$. This implies a photonic analog on of $2\xi^{-1}$. The trajectory Eqns.\ then read:
\begin{eqnarray}
\frac{b}{r_{\rm RF}\left(\phi\right)} &=& -\frac{r_C}{2b}\left[1+\cos\left(\phi\right)\right] + \sin\left(\phi\right),\\
\frac{b}{r\left(\phi\right)} &=& -\frac{1}{\xi b}\left[1+\cos\left(\phi\right)\right] + \sin\left(\phi\right).
\end{eqnarray}

\subsection{The hyperbolic sine case}
Now, lets assume a solution with the hyperbolic sine function:
\begin{equation}\label{eq:pathSmallSine}
r_{s}\left(\phi\right)=\frac{p_s}{e_s \sinh\left(\gamma_{s} \left[\phi-\phi_{0,s}\right]\right) + 1},
\end{equation}
again with $p_s=-p$ and $\gamma_s^2=-\gamma^2$. These orbits describe trajectories which approach from infinity but fall exponentially fast into the coordinate origin without a closest distance $r_m$. The unique infinite distance may be set to happen at $\phi=\pi$, giving the constant $\phi_{0,s}=\pi + \gamma_s^{-1}{\rm arcsinh}\left(1/e_s\right)$. The eccentricity may be obtained here by requiring the limit of $\lim_{\phi \to \pi} \sin\left(\phi\right) r_{s}\left(\phi\right) = b$ which yields an imaginary eccentricity $e_s=\pm i | b\xi |$, which would give imaginary radii. However, the hyperbolic sine gives trajectories which approach from infinity if one sets $\phi_0=0$. Then, the approach direction may still be specified by the choice of $e_{s}=-1/\sinh \left(\gamma \pi\right)$, however the distance to the optical axis is then fixed to be $\tilde{b}=p\gamma^{-1}\tanh\left(\pi\gamma\right)$ which does not reduce to $b$. The degree of freedom to achieve this was lost when the integration constant $\phi_{0,s}$ was discarded.
\subsection{Derivation of the differential scattering cross-section}
Using $\theta=2\phi_0-\pi$ and $\arccos\left(x\right)=\arctan\left(\sqrt{1-x^2}/x\right)$ for $x\ge 0$ we can write with Eqs.\ (\ref{eq:Parms1}), (\ref{eq:Parms2}):
\begin{equation}
\theta=\pi-2|\gamma^{-1}|\arctan\left(\gamma e\right)
\end{equation}
Since $\arctan\left(x\right)+\arctan\left(1/x\right)=\pi/2$ one has for $\gamma\approx 1$, i.e.\ the first order approximation, the following relation, $\theta\approx 2\arctan\left(1/e\right)$ or equivalently
\begin{equation}
\cot\left(\frac{\theta}{2}\right)=b\xi.\label{eqThetaCot}
\end{equation}
The geometric definition of the differential scattering cross-section is
\begin{equation}
2\pi \left(\frac{\mathrm{d}\sigma}{\mathrm{d}\Omega}\right)\sin\left(\theta\right)\mathrm{d}\theta=-2\pi b\mathrm{d}b.\label{eqndiffcross}
\end{equation}
From Eq.\ (\ref{eqThetaCot}) and the derivative $\cot'\left(x\right)=\sin^{-2}\left(x\right)$ we find:
\begin{equation}
b\mathrm{d}b=-\frac{\cot\left(\frac{\theta}{2}\right)}{\xi} \sin^{-2}\left(\frac{\theta}{2}\right)\frac{\mathrm{d}\theta}{2\xi}\label{eqnbdb}
\end{equation}
Combining Eqs.\ (\ref{eqnbdb}) and (\ref{eqndiffcross}), and using $\sin\left(2x\right)=2\sin\left(x\right)\cos\left(x\right)$ we write:
\begin{equation}
\frac{\mathrm{d}\sigma}{\mathrm{d}\Omega}=\frac{\cot\left(\frac{\theta}{2}\right)}{\sin\left(\theta\right)2\xi^2} \sin^{-2}\left(\frac{\theta}{2}\right)=\frac{1}{4\xi^2\sin^4\left(\frac{\theta}{2}\right)}
\end{equation}
Only the cross section for scattering at a greater angle than some chosen angle, $\sigma_{>\theta}$, is defined, and evaluates to, using a further trigonometric identity for $\cot\left(\frac{\theta}{2}\right)$ and $\int_{0}^{2\pi}\int_{\Theta}^{\pi}\sin^{-4}\left(\frac{\theta'}{2}\right)\sin\left(\theta'\right)\mathrm{d}\theta'\mathrm{d}\phi=4\pi\cot^2\left(\frac{\Theta}{2}\right)$:
\begin{equation}
\sigma_{>\Theta}\approx\frac{\pi}{\xi^2}\cot^2\left(\frac{\Theta}{2}\right)=\frac{\pi}{\xi^2}\left(\frac{1+\cos\left(\Theta\right)}{1-\cos\left(\Theta\right)}\right)
\end{equation}
\subsection{The GOA Shadow Region}
For the Coulomb scattering, one finds\cite{Warner1991,Adolph1972,Samengo1999}:
\begin{eqnarray}
r_{\rm RF,s}\left[1 - \cos\left(\phi\right)\right]&=& 2 r_C,\\
r_s\left[1 - \cos\left(\phi\right)\right]&=& 4\xi^{-1}.
\end{eqnarray}
In order to obtain the shadow region, one may consider for simplicity the intersection of neighboring asymptotes $y_A^b\left(z\right)$ and $y_A^{b+\mathrm{d}b}\left(z\right)$ and find the corresponding $z_i$-coordinate of intersection for $\mathrm{d}b\rightarrow 0$. There will be a solution for each $b$ that plays the role of a parameter describing the intersection-point curve $y_i\left(z\right)$ (green lines in the figure above). The Asymptote was Taylor-expanded to second order around $b\xi=\infty$ to obtain approximate results. Using $y_A\left(z\right)=b+\tan\left(\theta\right)\left[z+b\tan\left(\phi_0-\pi/2\right)\right]$ and requiring $\partial_b\, y_A^b\left(z_i\right)$ one finds
\begin{align}
z_i&\approx \frac{b^2\xi^2-2}{2\xi}\\
y_A^b\left(z_i\right)&\approx 2b-\frac{\pi}{4\xi}-\frac{\pi}{2b^2\xi^3}+\dots\\
y_i \left(z\right)\xi&\approx 2^{3/2}\sqrt{z\xi+1}-\frac{\pi\left(2+z\xi\right)}{4\left(1+z\xi\right)}\label{eqnShadowCurve}
\end{align}
such that $y_i \left(z\right)\approx 2^{3/2}\sqrt{z/\xi} - \xi^{-1}\pi/4$. This defines a shadow-region which spans an angle $\theta_s$ given by $\tan\left(\theta_s\right)=\lim_{z\rightarrow\infty} y_i\left(z\right)/z$. This angle goes to zero. More precisely, $\theta_s\approx 2^{3/2}\sqrt{\frac{1}{z\xi}}- \frac{\pi}{4}\frac{1}{z\xi}$.

\bibliographystyle{unsrt}
\bibliography{PRBib}

\end{document}